\documentclass[a4paper,UKenglish,cleveref, autoref, thm-restate]{oasics-v2021}

\hideOASIcs  

\title{BWT for string collections} 

\titlerunning{BWT for string collections} 

\author{Davide Cenzato}{Ca' Foscari University of Venice, Italy}{davide.cenzato@unive.it}{https://orcid.org/0000-0002-0098-3620}{}

\author{Zsuzsanna Lipt\'ak}{University of Verona, Italy}{zsuzsanna.liptak@univr.it}{https://orcid.org/0000-0002-3233-0691}{}

\author{Nadia Pisanti}{University of Pisa, Italy}{nadia.pisanti@unipi.it}{https://orcid.org/0000-0003-3915-7665}{}

\author{Giovanna Rosone}{University of Pisa, Italy}{giovanna.rosone@unipi.it}{https://orcid.org/0000-0001-5075-1214}{}

\author{Marinella Sciortino}{University of Palermo, Italy}{@unipa.it}{https://orcid.org/0000-0001-6928-0168}{}

\authorrunning{D.\ Cenzato et al.} 

\Copyright{Davide Cenzato and Zsuzsanna Lipt\'ak and Nadia Pisanti and Giovanna Rosone and Marinella Sciortino} 
\ccsdesc[500]{Theory of computation~Data structures design and analysis}

\keywords{Burrows-Wheeler transform, Extended Burrows-Wheeler transform, compressed text indexes, text compression, string collections, bioinformatics} 

\category{} 

\relatedversion{} 

\acknowledgements{DC is funded by ERC StG “REGINDEX: Compressed indexes for regular languages with applications to computational pan-genomics” grant nr 101039208. Views and opinions expressed are however those of the author(s) only and do not necessarily reflect those of the European Union or the European Research Council Executive Agency. Neither the European Union nor the granting authority can be held responsible for them. GR and NP are partially supported by PNRR - M4C2 - Investimento 1.5, Ecosistema dell'Innovazione ECS00000017 - ``THE - Tuscany Health Ecosystem'' - Spoke 6 ``Precision medicine \& personalized healthcare'', funded by the European Commission under the NextGeneration EU programme. MS is partially supported by Project ``ACoMPA'' (CUP B73C24001050001) funded by the NextGeneration EU programme PNRR MUR M4 C2 Inv. 1.5 -- Project ECS00000017 Tuscany Health Ecosystem (Spoke 6), CUP  B63C22000680007. ZsL, NP, GR, MS, are partially supported by MUR PRIN 2022 YRB97K ``PINC'' funded by Next Generation EU PNRR M4 C2 Inv. 1.1. NP is also supported by the PANGAIA and  ALPACA projects that received funding from the European Union’s Horizon 2020 research and innovation programme under the Marie Skłodowska-Curie grant agreements No.\ 872539 and 956229, respectively. ZsL, GR, and MS are also partially supported by the INdAM - GNCS Project CUP$\_$E53C24001950001. 
}

\nolinenumbers 

\EventEditors{John Q. Open and Joan R. Access}
\EventNoEds{2}
\EventLongTitle{42nd Conference on Very Important Topics (CVIT 2016)}
\EventShortTitle{CVIT 2016}
\EventAcronym{CVIT}
\EventYear{2016}
\EventDate{December 24--27, 2016}
\EventLocation{Little Whinging, United Kingdom}
\EventLogo{}
\SeriesVolume{42}
\ArticleNo{23}

\usepackage{todonotes}
\usepackage{xspace,xargs}
\usepackage{xcolor}
\usepackage{colortbl}

\usepackage{amsmath,amssymb,tabularx}
\usepackage{multirow}
\usepackage{enumerate}
\usepackage{url}
\usepackage{arydshln}
\usepackage{graphicx}
\usepackage{float}
\usepackage{tikz}
\usetikzlibrary{matrix}
\tikzset{every node/.append style={text depth=0.4ex}}
\usepackage[boxruled,vlined,linesnumbered]{algorithm2e}

\colorlet{darkred}{red!80!black}
\colorlet{darkblue}{blue!60!black}
\colorlet{darkgreen}{green!60!black}
\colorlet{medgreen}{green!70!black}
\colorlet{darkgray}{white!50!black}

\newcommand{\orange}[1]{{\color{orange} #1}}
\newcommand{\red}{\textcolor{red}}
\newcommand{\blue}{\textcolor{blue}}
\newcommand{\green}{\textcolor{medgreen}}

\newcommand{\violet}{\textcolor{violet}}

\def\rle(#1){\texttt{rle}(#1)}

\newcommand{\lex}{\ensuremath{\mathrm{lex}}}
\newcommand{\col}{\ensuremath{\mathrm{colex}}}
\newcommand{\rev}{\ensuremath{\mathrm{rev}}}
\newcommand{\same}{\;\hat{=}\;}
\newcommand{\opt}{\ensuremath{\mathrm{opt}}}

\def\bigM{\mathcal{M}\xspace}

\def\eBWT{\ensuremath{\textrm{EBWT}}}
\def\dollarebwt{\ensuremath{\textrm{dolEBWT}}}
\def\mdollar{\ensuremath{\textrm{mdolBWT}}}
\def\mdolBWT{\ensuremath{\textrm{mdolBWT}}}
\def\mdollarE{\ensuremath{\textrm{mdolEBWT}}}
\def\mdolEBWT{\ensuremath{\textrm{mdolEBWT}}}
\def\multiEBWT{\ensuremath{\textrm{mdolEBWT}}}

\def\colex{\ensuremath{\textrm{colexBWT}}}
\def\optimal{\ensuremath{\textrm{optBWT}}}
\def\plus{\ensuremath{\textrm{plusBWT}}}
\def\concat{\ensuremath{\textrm{concatBWT}}}

\newcommand{\EBWT}{\ensuremath{\mathrm{EBWT}}\xspace}
\newcommand{\BWT}{\ensuremath{\mathrm{BWT}}\xspace}
\newcommand{\bwt}{\ensuremath{\mathrm{BWT}}\xspace}

\newcommand{\GSA}{\ensuremath{\mathrm{GSA}}\xspace}
\newcommand{\GCA}{\ensuremath{\mathrm{GCA}}\xspace}
\newcommand{\CA}{\ensuremath{\mathrm{CA}}\xspace}
\newcommand{\SA}{\ensuremath{\mathrm{SA}}}

\newcommand{\conj}{\mathrm{conj}}
\renewcommand{\alph}{\ensuremath{\mathrm{alph}}}
\renewcommand{\root}{\ensuremath{\mathrm{root}}}
\renewcommand{\exp}{\ensuremath{\mathrm{exp}}}
\newcommand{\cais}{\ensuremath{\mathtt{cais}}}

\begin{document}

\maketitle

\begin{abstract}
We survey the different methods used for extending the BWT to collections of strings, following largely [Cenzato and Lipták, CPM 2022, Bioinformatics 2024]. We analyze the specific aspects and combinatorial properties of the resulting BWT variants and give a categorization of publicly available tools for computing the BWT of string collections. We show how the specific method used impacts on the resulting transform, including the number of runs, and on the dynamicity of the transform with respect to adding or removing strings from the collection. 

We then focus on the number of runs of these BWT variants and present the optimal BWT introduced in [Cenzato et al., DCC 2023], which implements an algorithm originally proposed by [Bentley et al., ESA 2020] to minimize the number of BWT-runs. We also discuss several recent heuristics and study their impact on the compression of biological sequences. 

We conclude with an overview of the applications and the impact of the BWT of string collections in bioinformatics.
\end{abstract}

\bigskip\bigskip

\newpage 


\section{Introduction}

With the advent of high-throughput sequencing technologies, the amount of genomic data available in public databases has undergone an exponential increase~\cite{Stephens15}. Similar data explosion can be witnessed in other areas based on textual data, such as webpages, versioned texts (such as Wikipedia), versioned software repositories (such as Github), digital books, and many others (for more details, see~\cite{Navarro21a}). 

However, not only has the amount or size of textual data increased, but it has also changed character. Most importantly, nowadays most data of interest consists of {\em string collections} rather than of individual strings. For the purpose of this survey, a string collection is any multiset of strings (or sequences, two terms we use interchangeably). 
Early examples include the famous 1000 Genomes Project~\cite{1000genomes}, in the course of which almost $2000$ human genomes were sequenced, 
followed by the 10,000 Genomes Project~\cite{10K}, the 100,000 Human Genome Project~\cite{100K}, the 1001 Arabidopsis Project~\cite{arab}, the 3,000 Rice Genomes Project  (3K RGP)~\cite{rice}, and others. Projects of the last few years include the COVID-19 Genomics UK Consortium's~\cite{COVID-19-consortium} and the H3Africa Consortium's work~\cite{H3Africa}. 

The Burrows-Wheeler Transform (BWT)~\cite{bwt94} is one of the most fundamental methods for string processing and compressed data storage~\cite{fm-index-jacm,MakinenN05,r-index-jacm}; the tools built upon these BWT-based compressed data structures are among the most used in bionformatics~\cite{bowtie,bwa,bowtie2,soap}. 
It is therefore natural that researchers and practicioners have gone ahead and applied BWT-based methods to this new kind of data. The BWT, however, was originally defined for {\em one} string, and it is not immediately clear how to extend it to several strings. 

\medskip

As it turns out, several different methods have been used to generalize the BWT to string collections. Some of these methods were carefully designed and consciously chosen, while others appear to have been simply implementation choices. In fact, in many publications it is not explicitly specified how the BWT of a string collection was computed. Underlying this is the implicit assumption---incorrect, as we will see---that all methods are equivalent. 

The classic method, paralleling {\em generalized suffix trees} and {\em generalized suffix arrays} (see e.g., ~\cite{Gusfield1997} or~\cite{Ohlebusch2013}), is to concatenate the strings in the collection, using distinct end-of-string symbols to separate consecutive strings. In effect, this means turning the string collection into one string, at which point one can apply the classic BWT in the usual way. Following established terminology, we will refer to these end-of-string-symbols as 'dollars'; so we have a distinct dollar ($\mathtt{\$}_i$) for each input string. In actual implementations, of course, the same dollar-symbol is used, because otherwise the alphabet would be blown up enormously, not a viable choice.  However, the dollars remain distinct conceptually; one common way to handle this is to use the position of the dollars for breaking ties. This method, concatenating the strings with distinct dollars, is what we will refer to  in this paper as {\em multidollarBWT}. 

Another method that is commonly used is to concatenate the input strings but drop the distinction between the dollars; this is a much simpler choice on an implementation level because one does not have to take care of distinguishing between the different dollars. In order to ensure correctness (of BWT construction, LF-mapping, backward search, etc.), it is necessary to append one more symbol at the very end of the concatenated string (often termed 'hash' and denoted $\mathtt{\#}$), which is smaller than all other characters, including the dollar. We will call this method {\em concatBWT}. 

In 2007, a mathematically clean extension of the BWT to string collections was proposed by Mantaci et al.~\cite{MantaciRRS2007_ebwt}, which the authors called {\em extended BWT} (\EBWT). The \EBWT is essentially the inverse of the Gessel-Reutenauer bijection~\cite{GesselR93}, known in combinatorics on words, but hitherto unknown in the research community working on textual data structures and string processing. The \EBWT is a generalization of the BWT from one string to a multiset of primitive strings. Similarly to the BWT, it is a permutation of the characters of the strings in input and therefore has length equal to the total length of the multiset in input. It also allows backward search and has the same 'clustering effect' as the BWT, i.e., it tends to produce long same-character runs on repetitive inputs. We refer to this method simply as {\em \EBWT}. 

Applying the classic BWT to a string which terminates with a dollar is different in many ways from applying it to one without. (For example, $\bwt(\mathtt{banana\$}) = \mathtt{annb\$aa}$, while $\bwt(\mathtt{banana}) = \mathtt{nnbaaa}$.) Similarly, appending a dollar to the ends of the input strings changes the properties of the \EBWT. Again, one possibility is to append the same dollar to every input string; this is what happens, for example, reading in strings line-by-line, since each line is terminated by a newline symbol. We will call this method {\em dollarEBWT}. 

Finally, if one appends different dollars to the input strings (or the same dollar but considering them as different), then we get what we call {\em multidollarEBWT}. Regarding the resulting transform, this method is similar to the multidollarBWT method (concatenate the strings with distinct dollars separating them), except that the indices of the dollars in the final transform are shifted by one. 


In Table 1, we give a classification of existing BWT construction tools, geared specifically to string collections, according to the BWT variant they produce. As it is not always possible to distinguish the underlying method, our categorization is based on a combination of the information given in the accompanying papers and on reverse engineering the final output of the tools.

\begin{table*}[t]
\centering
\renewcommand{\arraystretch}{1.2}
\scalebox{0.86}{
\begin{tabular}{l| l| l }
 & {\em result on example} & {\em tools} \\
\hline\hline
\multicolumn{3}{l}{\textit{BWT variants} (Sections~\ref{sec:BWTvariants} and \ref{sec:properties})}\\
\hline 
\eBWT & {\tt GGGCTACTCACACCTCTAGCG} & {\tt PFP-eBWT}~\cite{pfp-ebwt-tool-BCLRS_SPIRE2021_ebwt}, {\tt cais}~\cite{cais-tool-BCLRS_SPIRE2021_ebwt}, {\tt lFGSACA}~\cite{lfgsaca-tool-Olbrich_lfgsaca} \\
\hline
\dollarebwt & {\tt ACACAGGGCGCCTAT\$\$\$TCTC\$\$G\$C} & {\tt G2BWT}~\cite{bwt-imp-tool-DominguezN21}, {\tt PFP-eBWT}~\cite{pfp-ebwt-tool-BCLRS_SPIRE2021_ebwt}, {\tt cais}~\cite{cais-tool-BCLRS_SPIRE2021_ebwt}, {\tt msbwt}~\cite{msbwt-tool-HoltMcMillan2014}\\
\hline
 \multirow{4}{*}{\shortstack{\mdolEBWT \\[1ex] and \\[1ex] \mdolBWT}} & \multirow{4}{*}{\tt AGCACAGCGGCCTTA\$\$\$TTCC\$\$G\$C} & {\tt BEETL}~\cite{beetl-tool-BauerCoxRosoneTCS2013}, {\tt BCR\_LCP\_GSA}~\cite{BCR-tool-BauerCoxRosoneTCS2013}, {\tt ropeBWT2}~\cite{ropebwt2-tool-Li14a}, \\
  &  & {\tt ropeBWT3}~\cite{ropebwt3-tool-ropebwt3}, {\tt Merge-BWT}~\cite{merge-tool-Siren16}, \\
  && {\tt gsufsort}~\cite{gsufsort-tool-LTGPR2020gsufsort}, {\tt grlBWT}~\cite{grlBWT-tool-DiazNavarro2023_grlBWT}, {\tt eGAP}~\cite{egap-tool-EgidiLMT19_egap}, {\tt eGSA}~\cite{egsa-tool-Louza2017d}, \\
 &&{\tt bwt-lcp-parallel}~\cite{bwt-lcp-parallel-tool-BonizzoniVPPR19}, {\tt bwt-lcp-em}~\cite{bwt-lcp-em-tool-BonizzoniDVPPR_2021}\\
\hline 
\multirow{2}{*}{\concat} & \multirow{2}{*}{\tt  \$ACAGCAGCGGCCTAT\$\$\#TCTC\$\$G\$C} & {\tt BigBWT}~\cite{big-bwt-tool-BGKLMM2019_bigBWT}, {\tt r-pfbwt}~\cite{r-pfbwt-tool-OlivaGB23}, {\tt CMS-BWT}~\cite{CMS-BWT-tool-Masillo23},\\ && tools for single-string BWT\\
\hline \hline
\multicolumn{3}{l}{\textit{specific string orders} (Section~\ref{sec:opt})}\\
\hline 
colexBWT & {\tt AAACCGCGGGCCTAT\$\$\$TCTC\$\$G\$C} & {\tt BCR\_LCP\_GSA}~\cite{BCR-tool-BauerCoxRosoneTCS2013}, {\tt ropeBWT2}~\cite{ropebwt2-tool-Li14a}, {\tt ropeBWT3}~\cite{ropebwt3-tool-ropebwt3}\\ 
\hline 
plusBWT & {\tt AAACCGGGCGCCTTA\$\$\$TTCC\$\$G\$C} & {\tt BCR\_LCP\_GSA}~\cite{BCR-tool-BauerCoxRosoneTCS2013} \\
\hline 
\optimal & {\tt AAAGCCCGGGCCTTA\$\$\$TTCC\$\$G\$C} & {\tt optimalBWT}~\cite{optimalBWT-tool-CGLR_DCC2023_opt}\\
\hline 
\end{tabular}
}
\vspace{2mm}
\caption{\label{tab:tools}The five \BWT\ variants and the three \BWT\!\!\! s using specific string orders, on the multiset ${\cal M} = ( {\tt CTGA, TG, GTCC, TCA, CGACC, CGA})$. 
Double listings due to options of the software. We note also that {\tt lFGSACA} is a library, while the others are dedicated command-line tools for BWT computation. 
}
\end{table*}

In this paper, we survey the five different methods sketched above, giving their exact definitions and combinatorial properties. As we will see, which method is chosen has important consequences for the resulting transform. This includes issues such as input order dependence (giving the string collection in a different order may or may not result in different outputs), dynamicity (how does adding or removing a string from the collection impact on the transform), and number of runs of the BWT. This last parameter, often denoted $r$, is central in BWT-based data structure design, since modern compressed data structures such as the $r$-index~\cite{r-index-jacm} require storage space linear in $r$. 

We will in fact show that if the goal is to reduce the number of runs, then either the  multidollarEBWT or the multidollarBWT should be chosen, because methods leading to few runs---the exact method, called {\em optimalBWT}, or heuristics like {\em colexBWT} or {\em plusBWT}---only work with these variants. Moreover, both allow adding or removing strings from the collection without having to recompute the BWT. On the other hand, if input order independence is desired, then the \EBWT 
or the dollarEBWT are the right choice. 

We will also see that the concatBWT, the variant where the strings are concatenated using the same dollar, has several undesirable properties. 

The number $r$ of runs of the BWT appears increasingly as a parameter characterizing the input data: $r$, or equivalently, the average runlength $n/r$ of the BWT, is often used as a measure of how repetitive the data is. This is well justified in the case of one string, as many recent publications on the theory of $r$ as a repetitiveness measure attest (see~\cite{Navarro21a} for a recent survey). However, for string collections, the parameter $r$ is not well defined, given that different methods on the same input collection can give different $r$'s, and even the same method on the same collection, if presented in a different order, can give very different $r$'s (up to a multiplicative factor of over $31$ on real data~\cite{CGLR_DCC2023_opt}). We argue that this parameter should be standardized, to $r_{\opt}$, the minimum number of runs among all possible transforms on a given string collection. 


We round off the paper with a section about applications in bionformatics that make use of string collections. These are highly repetitive data sets that are the ideal object for data structures based on the BWT. We give an overview of the different areas and problems using highly repetitive sequence collections, which are at the center of today's bioinformatics research.

\subsection{Overview of paper}

The paper is organized as follows. We provide necessary background and terminology in \cref{sec:basics}. In \cref{sec:BWTvariants}, we give the precise definitions and examples for the five BWT variants, distinguishing between \EBWT-based variants (\cref{sec:ebwt-based}) and concatenation-based variants (\cref{sec:concatenation-based}). We study their combinatorial properties in \cref{sec:properties}, followed by a section on reducing the number of runs of the BWT (\cref{sec:opt}). This section also contains experimental results on real data, showing how much the number of runs can be reduced. \cref{sec:bioinformatics} contains an overview of areas in bioinformatics where string collections play an important role. We close with a summary and some suggestions in \cref{sec:conclusion}. 

The classification of BWT variants for string collections presented in this paper is based largely on that contained in~\cite{CenzatoL22,CenzatoLiptak_bioinformatics2024} but with some additions and modifications. In particular, \cref{sec:BWTvariants}, containing the definitions, has been extended, and the classification has been further developed\footnote{We now group the BWT variants in \EBWT-based versus concatenation-based, and we subsume colexBWT and optBWT under multidollarEBWT and multidollarBWT.}. All examples contained in this paper are new. The combinatorial properties of the BWT variants of \cref{sec:properties} are partially from~\cite{CenzatoL22,CenzatoLiptak_bioinformatics2024}, with updated presentation; \cref{sec:dynamicity}, on dynamicity of the BWT variants, has been newly added and \cref{sec:output_order}, on the output order of the BWT variants, generalizes results presented in~\cite{CenzatoL22}. Finally, \cref{sec:opt} contains mostly results from~\cite{CGLR_DCC2023_opt,BCGR_SAP_CPM2024}, while the survey of bioinformatics applications contained in Section~\ref{sec:bioinformatics} is not contained elsewhere.


\section{Basics} \label{sec:basics}

Let $\Sigma =\{a_1, a_2, \ldots, a_\sigma\}$ be a finite ordered alphabet $\Sigma$ with $a_1< a_2< \ldots < a_\sigma$. 
A string (or text) $T=T[1..n]$ is a sequence of symbols $T[1]\cdots T[n]$ over the alphabet $\Sigma$. We denote  by $\epsilon$ the empty string and by $|T|$ the length of string $T$. For $1\leq i\leq n$, the substring $T[1..i]$ is called the $i$th {\it prefix} of $T$ and $T[i..n]$ the $i$th {\it suffix} of $T$. The string ${T}^{\rev}=T[n]\cdots T[1]$ is the \emph{reverse} string of $T$. Let $<_\lex$ denote the standard lexicographic order. The \emph{colexicographic} order (sometimes also referred to as \emph{reverse lexicographic order}) is defined as $S<_{\col} T$ if and only if $S^{\rev}<_{\lex} T^{\rev}$.

Given a string $T$, the string $T^k=TT \cdots T$ is the string obtained by concatenating $k$ copies of $T$. A string $T$ is called \emph{primitive} if $T=U^k$ implies $T=U$ and $k=1$. For every string $T$, there exists a unique primitive string $U$ and a unique integer $k$ such that $T=U^k$. The string $U$ is called $\root(T)$ and $k$ is called $\exp(T)$; thus, $T=\root(T)^{\exp(T)}$. 

For a string $T$, we denote by $T^\omega=TT\cdots$ the infinite string obtained by concatenating an infinite number of copies of $T$. 
The {\em omega-order} on $\Sigma^*$ is defined as follows: $S \prec_\omega T$ if (a) $S^\omega <_\lex T^\omega$, or (b) $S^\omega = T^\omega$ and $\exp(S)<\exp(T)$ (note that $S^\omega = T^\omega$ implies $\root(S)=\root(T)$). It can be verified that the $\omega$-order relation coincides with the lexicographic order if neither of the two strings is a proper prefix of the other. Otherwise, the two orders can be different. For instance,  ${\tt CGA <_\lex CGACC}$ but ${\tt CGACC\prec_\omega CGA}$.

Given string $T$, the {\em alphabet of $T$} is defined as $\alph(T) = \{ T_i \mid 1\leq i \leq |T|\}$. 
A \emph{run} in a string $T$ is a maximal substring $U$ such that $\alph(U)$ is a singleton. Given a string $T$, we denote by $\rho(T)$ the number of runs of $T$, e.g., $\rho(\mathtt{baaabbc})=4$. Moreover, we denote by $r(T)$ the number of runs of the BWT of $T$ (defined below).

The string $S$ is a \emph{conjugate} (or {\em cyclic rotation}, or simply {\em rotation}) of the string $T$ if $S=T[i..n]T[1..i-1]$, for some $i \in \{1,\ldots, n\}$. In this case, $S$ is also called the $i$th conjugate of $T$, denoted $\conj_i(T)$. %
The \emph{conjugate array} $\CA$ of a string $T$ of length $n$ is the permutation of $\{1,\ldots,n\}$ such that $\CA[j]=i$ if $\conj_i(T)$ is the $j$th conjugate of $T$ with respect to the lexicographic order, with ties broken according to string order, i.e.\ if $\CA[j] = i$ and $\CA[j'] = i'$ for some $j<j'$, then either $\conj_i(T) <_\lex \conj_{i'}(T)$, or $\conj_i(T) = \conj_{i'}(T)$ and $i<i'$. If $T$ is a primitive string, a conjugate $S$ of $T$ is called \emph{Lyndon rotation} of $T$, denoted as $Lynd(T)$, if it is lexicographically the smallest among all conjugates of $T$.

To define the {\em suffix array} $\SA$, it is common to assume that the string has an end-of-string symbol $\$$ appended at the end, such that $\$$ is smaller than all other characters and does not occur elsewhere in the string. Then, the $\SA$ is a permutation of the indices $\{1,2,\ldots,n+1\}$ such that $\SA[j]=i$ if suffix $T[i..n+1]$ is the $j$th suffix of all suffixes in lexicographic order. Note that in this case, there can be no ties, since two distinct suffixes necessarily have different lengths. Since the string $T\$$ has exactly one occurrence of $\$$, the conjugate array and the suffix array of $T\$$ coincide. 

The \emph{Burrows-Wheeler Transform} (\BWT)~\cite{bwt94} is a reversible transformation that permutes the characters of the input string. Given a string $T$, the {\em BW-matrix} of $T$ is defined as the matrix whose rows are the conjugates of $T$, given in lexicographic order. 
We denote by $\bwt(T)$ the last column of the matrix. The transformation also outputs an integer $i$, which is the position of 
$T$ within the matrix.
It follows from the definition that if a string $S$ is a conjugate of $T$, then $\bwt(S)=\bwt(T)$. The index $i$ is fundamental for recovering $T$ from $\bwt(T)$, otherwise the original string can be recovered only up to conjugates. When the reconstruction of the original string is not relevant, the index is omitted.

The \emph{standard permutation}  of a word $S=S[1..n]$ over $\Sigma$ is the permutation $\pi_S$ of $\{1,2,\ldots,n\}$ such that $\pi_S(i)<\pi_S(j)$ if and only if $S[i]<S[j]$ or  $S[i]=S[j]$ and $i<j$. The standard permutation $\pi_S$ of a string $S$ is also called LF-mapping if $S=\bwt(T)$, for some string $T$. It is known that $T$ is a primitive word if and only if $\pi_{S}$ is a cycle of length $n$. 
In this case, given $S$, $\pi_S$ and an index $i$, the $i$th conjugate $U$ of $T$ in lexicographic order can be recovered as follows: $U[n]=S[i]$ and $U[n-j]=S[\pi^j_S(i)]$, for $j=1,\ldots,n-1$.
The conjugate $Lynd(T)$ can be recovered with $i=1$. 

The permutation $\bwt(T)$ can be computed from the conjugate array $\CA$, since for all $j=1,\ldots, n$, $\bwt(T)[j]=T[\CA[j] - 1]$, assuming that $T[0]=T[n]$. 
Alternatively, if the input string $T$ terminates with an end-of-string symbol $\$$, $\bwt(T\$)$ can be computed from the suffix array $\SA$. Moreover, recovering the first row of the BW-matrix will yield $\$T$, because $\$$ is smaller than all other symbols, and since $\$$ occurs exactly once, unique recovery is ensured also in this case. 

Appending an end-of-string symbol to the string $T$ has traditionally allowed $\bwt(T\$)$ to be computed using linear-time algorithms for suffix array construction. However, it is possible to compute $\bwt(T)$ in linear time without appending any end-of-string symbol $\$$ to $T$. This can be done by computing the suffix array of $U=Lynd(\root(T))$~\cite{GIA07} and then obtaining $\bwt(T) = \bwt(U^k)$, where $k=\exp(T)$, from $\bwt(U)$, applying a simple relationship between the two transforms: $\bwt(U^k)$ can be obtained by concatenating $k$ copies of each character of $\bwt(U)$~\cite{MantaciRS03}. 
Alternatively, it is also possible to compute $\bwt(T)$ directly in linear time with the $\cais$ algorithm, which is based on induced sorting \cite{BCLRS_SPIRE2021_ebwt}.   

Now let ${\cal M} = \{T_1, T_2, \ldots, T_m\}$ be a multiset of strings (often referred to as a {\em string collection}), with $|T_d|=n_d$ for $1\leq d \leq m$. We denote by $|{\cal M}| = m$ the number of strings in ${\cal M}$ and by $||{\cal M}|| = \sum_{d=1}^m n_d$ their total length. Further, we will denote by ${\cal M}_{\$} = \{T_1\$_1, T_2\$_2, \ldots, T_m\$_m\}$ the set of strings with end-of-string markers appended to all strings in ${\cal M}$, of total length $m + \sum_{d=1}^m n_d$. 

The \emph{generalized conjugate array} $\GCA$ of the collection ${\cal M}=\{T_1, T_2, \ldots, T_m\}$ is an array of total length $||{\cal M}||$, 
where $GCA[j] = (d, i)$ if $conj_i(T_d)$ is the $j$th string in the $\omega$-sorted list of the conjugates of the strings in ${\cal M}$, with ties broken first w.r.t. the index of the string (in case of identical strings), and then w.r.t.\ the position in the string (in case of identical conjugates of the same string). 

The \emph{generalized suffix array} $\GSA$ of the collection ${\cal M}_{\$}$ is an array of length $||{\cal M}_\$|| = m + \sum_{d=1}^m n_d$,
containing pairs of integers $(d,i)$, corresponding to the lexicographically sorted suffixes of the strings in ${\cal M}_\$$. 
In particular, $\GSA[j]=(d,i)$ is the pair corresponding to the $j$th smallest suffix of the strings in ${\cal M}_{\$}$, with ties broken according to $\$_1 < \$_2 < \ldots < \$_m$.

\section{BWT variants for string collections}\label{sec:BWTvariants}

For string collections, several variants of the BWT have been proposed. Each of these variants aims to extend the features of the BWT for single strings to collections of strings. In this section, we review the most significant variants, which form the foundation of the most widely used 
tools for indexing and compressing string collections based on the BWT. We use the classification introduced in \cite{CenzatoL22,CenzatoLiptak_bioinformatics2024}, and in addition group these variants into two families, depending on the strategy used to process the strings of the collection.\footnote{In difference to~\cite{CenzatoL22,CenzatoLiptak_bioinformatics2024}, here we include a new separator-based BWT variant, the multidollarEBWT, which was previously subsumed under multidollarBWT.} 
The first family includes transformations that explicitly or implicitly apply the Extended BWT ($\EBWT$) introduced in~\cite{MantaciRRS2007_ebwt} to the multiset of strings, to which end-of-string symbols may or may not have been appended. The second family, on the other hand, consists of those transformations that involve concatenating the strings of the collection, and then applying the traditional $\BWT$ to the concatenated string. We will see that these BWT variants, while preserving the general functionality of the $\BWT$, can produce significantly different outputs. 

Let us denote by ${\cal M}=\{T_1, T_2, \ldots, T_m\}$ the input string collection, i.e.,  
a multiset of $m$ strings, with $|T_d|=n_d$ for $1\leq d \leq m$. Even though the input is a {\em multiset} of strings, in real applications, this input is always given in some order. As we will see, the input order plays an important role in the resulting transform for several of the BWT variants we will review. 
For this reason, in slight abuse of notation, whenever the definition depends on the input order, we will treat ${\cal M}$ in the following as an ordered collection and will write accordingly ${\cal M}=(T_1, T_2, \ldots, T_m)$. When it does not play a role, in the sense that the output is always the same whatever the input order, in those cases we will stick to the multiset notation. 

In Table \ref{tab:tools} we list, for each BWT variant, the tools that compute it (up to renaming of dollars).

\subsection{EBWT-based transforms}\label{sec:ebwt-based}

In this section, we present the three \EBWT-based transforms: \EBWT, \dollarebwt, and \mdolEBWT. See \cref{tab:eBWT-based-running-example} for an example. 

\subparagraph{Extended BWT}
The {\em Extended Burrows--Wheeler Transform}~\cite{MantaciRRS2007_ebwt} (\EBWT) is a generalization of the \BWT\ to a multiset of strings. Given the collection of strings ${\cal M}=\{ T_1, \ldots, T_m\}$, $\eBWT({\cal M})$ is a permutation of the characters of the strings in ${\cal M}$, obtained by concatenating the last characters of all conjugates of the strings of ${\cal M}$, listed according to the $\omega$-order. 
Note that $\EBWT$ is a transform on the string collections treated as multisets. Indeed, the output $\EBWT({\cal M})$ is independent of the order in which the strings appear in the collection ${\cal M}$. For this reason, we denote its input as a multiset.

Similarly to $\BWT$, the permutation $\eBWT(\mathcal{M})$ can be computed from the generalized conjugate array of $\mathcal{M}$ in linear time, since 
\begin{equation}
\mathsf {\eBWT}(\mathcal{M})[i] = {\left\{ 
\begin{array}{ll} 
   T_d[j-1] &  \text{ if } \GCA[i]=(d,j) \text{ with } j>1,\\ 
   T_d[n_d]       & \text{ if }  \GCA[i]=(d,j) \text{ with } j=1.
\end{array}
\right. } 
\end{equation}

The transformation also outputs the $m$-tuple $I=(i_1, \ldots, i_m)$ listing the positions 
of the strings of the collection within this $\omega$-sorted list. 
The $m$-tuple $I=(i_1, \ldots, i_m)$ is used to recover the multiset ${\cal M}$ from $\eBWT({\cal M})$. In fact, if $\pi_U$ is the standard permutation of the word $U=\eBWT({\cal M})$, 
then $\pi_U$ decomposes into $m$ disjoint cycles $\pi_1,\pi_2, \ldots, \pi_m$ such that the index $i_d$ appears in the cycle $\pi_d$. The strings of the collection $\mathcal{M}$ are recovered as follows: for every $d=1,\ldots,m$, $T_{d}[l_d-t]=\eBWT[{\pi_d}^t[i_d]]$, where $l_d$ is the length of the cycle $\pi_d$ and $t$ ranges from $0$ to $l_d-1$. Moreover, if the set of indices is considered in increasing order, then the strings of the collection are recovered in $\omega$-order. When the reconstruction of the original multiset is not relevant, the $m$-tuple is omitted.

\begin{example}\label{ex:ebwt}
Given the collection ${\cal M} = \{ {\tt \texttt{CTGA},\texttt{TG},\texttt{GTCC},\texttt{TCA},\texttt{CGACC}, \texttt{CGA} }\}$, the output of $\EBWT$ is the string $\texttt{GGGCTACTCACACCTCTAGCG}$ together with the indices $(9, 10, 12, 16, 18, 21)$, as shown in Table~\ref{tab:eBWT-based-running-example}. 
The LF-mapping decomposes into the cycles $(1\  13\  9\  7\  6)$ $(2\  14\  10)$ $(3\ 15\ 20\ 12)$ $(4\ 5\ 18)$ $(8\ 19\ 16\ 11)$ and $(17\ 21)$. Starting from the indices in the set $9$, $10$, $12$, $16$, $18$ and $21$, the strings $\texttt{CGACC}$, $\texttt{CGA}$, $\texttt{CTGA}$, $\texttt{GTCC}$, $\texttt{TCA}$ and $\texttt{TG}$ are reconstructed.
\end{example}

\subparagraph{Dollar-EBWT} This transformation, here denoted \dollarebwt, 
uses a single end-of-string symbol appended to each string in the collection. More precisely, given the collection ${\cal M}=\{T_1, \ldots, T_m\}$, 
\[\dollarebwt(\mathcal{M})=\eBWT(\{T_d\$ \mid d=1,\ldots,m\}).\]

Note that in this case, to recover the set of strings in the collection, it is sufficient to know the positions of the $\$$-symbols in $\dollarebwt(\mathcal{M})$. 
Since none of the strings in the collection is a prefix of another, the $\omega$-order coincides with the lexicographic order. For this reason, starting from the positions of the $\$$ symbols in increasing order, the strings of $\mathcal{M}$ can be recovered in lexicographic order.

Similarly to $\EBWT$, the permutation $\dollarebwt(\mathcal{M})$ can be computed from the generalized conjugate array $\GCA$ of the multiset $\{T_d\$ \mid T_d \in \mathcal{M}\}$ in linear time, since 
\begin{equation}
\dollarebwt(\mathcal{M})[i] = {\left\{ 
\begin{array}{ll} 
   T_d[j-1] &  \text{ if } \GCA[i]=(d,j) \text{ with } j>1,\\ 
   \$       & \text{ if }  \GCA[i]=(d,j) \text{ with } j=1.
\end{array}
\right. }
\end{equation}

\begin{example}
Let us consider the collection ${\cal M} = \{ {\tt \texttt{CTGA},\texttt{TG},\texttt{GTCC},\texttt{TCA},\texttt{CGACC}, \texttt{CGA} }\}$,
then $\dollarebwt({\cal M})= \texttt{ACACAGGGCGCCTAT\$\$\$TCTC\$\$G\$C}$, as shown in Table \ref{tab:eBWT-based-running-example}. 
The LF-mapping decomposes into $6$ cycles
$(1\ 7\  20\ 16)$ $(2\ 11\ 14\ 10\ 22\ 17)$ $(3\ 8\ 21\ 27\ 18)$ $(4\ 12\ 15\ 25\ 23)$ $(5\ 9\ 13\ 24)$ and $(6\ 19\ 26)$.
\end{example}

\subparagraph{Multidollar-EBWT} 

This transformation, here denoted $\multiEBWT$, 
uses distinct end-of-string symbols appended to the strings in the collection. Given the collection ${\cal M}=(T_1, \ldots, T_m)$ and the end-of-strings symbols $\$_1<\$_2<\cdots<\$_m$, 
\[\multiEBWT({\cal M})=\eBWT(\{T_d\$_d \mid d=1,\ldots,m\}).\]

Note that, unlike the previous transformations, the output of $\multiEBWT(\mathcal{M})$ is influenced by the order in which the strings appear in the collection, because the distinct end-of-string symbols result in different orders among the strings in the multiset. (For more on this, see \cref{sec:BWTvariants}.) 
Further, the position of each $\$_d$ allows for the reconstruction of the corresponding string $T_d$ of the collection. 

Since end-marker symbols are appended to each string in the collection, $\multiEBWT(\bigM)$ can be constructed using the lexicographic order of the suffixes of the strings in $\bigM$  as follows:  
\begin{equation} \label{eq.multiEBWT}
\multiEBWT(\bigM)[i] = {\left\{ 
\begin{array}{ll} 
    T_d\$_d[(j-1)] &  \text{ if } \GSA[i]=(d,j) \text{ with } j>1\\ 
    \$_d       & \text{ if }  \GSA[i]=(d,j) \text{ with } j=1
\end{array}
\right. } 
\end{equation}

\begin{example}
Let us consider the collection ${\cal M} = ( \mathtt{CTGA},\mathtt{TG},\mathtt{GTCC},\mathtt{TCA},\mathtt{CGACC}, \mathtt{CGA})$. Then 
$\multiEBWT(\mathcal{M})=\mathtt{AGCACAGCGGCCTTA\$_6\$_5\$_1TTCC\$_3\$_4G\$_2C }$, as shown in Table \ref{tab:eBWT-based-running-example}.
The LF-mapping decomposes into $6$ cycles 
$(1\ 7\  20\ 27\ 18)$ $(2\ 19\ 26)$ $(3\ 11\ 14\ 25\ 23)$ $(4\ 8\ 13\ 24)$ $(5\ 12\ 15\ 10\ 22\ 17)$ and $(6\ 9\ 21\ 16)$. 
For $\mathcal{M'}=(\mathtt{TCA}, \mathtt{CTGA},\mathtt{CGA},\mathtt{TG},\mathtt{GTCC},\mathtt{CGACC})$, where the same strings are given in a different order, $\multiEBWT({\cal M'})$ $=$ $\mathtt{AAAGCCCGGGCCTTA\$_3\$_6\$_2TTCC\$_5\$_1G\$_4C}$.  For more details, see Section~\ref{sec:opt}. 
\label{ex:ebwtOne}
\end{example}

\definecolor{customgray}{rgb}{0.60, 0.60, 0.60}
\definecolor{customgreen}{rgb}{0.0, 0.5, 0.0}
\newcommand{\cgray}[1]{{\textcolor{customgray}{ #1}}}
\newcommand{\cgreen}[1]{{\textcolor{customgreen}{#1}}}

\begin{table*}[h]
\centering
\ttfamily

\renewcommand{\arraystretch}{1.1}

\scalebox{0.78}{
\begin{minipage}[t]{4.0cm}
\vspace{-75.20mm}
\begin{tabular}{ |r|r|l|c| }
\hline
&\multicolumn{3}{c|}{\normalfont \eBWT}\\
\hline
\normalfont $i$&\normalfont GCA & \normalfont rotation &  \\
\hline
\normalfont1&\normalfont (5,3) & ACCCG & G \\
\normalfont2&\normalfont (6,3) & ACG   & \large \red{G} \\
\normalfont3&\normalfont (1,4) & ACTG  & \large \red{G} \\
\normalfont4&\normalfont (4,3) & ATC   & \large \red{C} \\
\normalfont5&\normalfont (4,2) & CAT   & T \\
\normalfont6&\normalfont (5,4) & CCCGA & \large \orange{A} \\
\normalfont7&\normalfont (5,5) & CCGAC & \large \blue{C} \\
\normalfont8&\normalfont (3,3) & CCGT  & \large \orange{T} \\
\normalfont9&\normalfont (5,1) & CGACC & \large \cgreen{C} \\
\normalfont10&\normalfont (6,1) & CGA   & \large \cgreen{A} \\
\normalfont11&\normalfont (3,4) & CGTC  & \large \blue{C} \\
\normalfont12&\normalfont (1,1) & CTGA  & \large \cgreen{A} \\
\normalfont13&\normalfont (5,2) & GACCC & C \\
\normalfont14&\normalfont (6,2) & GAC   & \large \violet{C} \\
\normalfont15&\normalfont  (1,3) & GACT  & \large \violet{T} \\
\normalfont16&\normalfont (3,1) & GTCC  & \large \cgreen{C} \\
\normalfont17&\normalfont  (2,2) & GT    & T \\
\normalfont18&\normalfont (4,1) & TCA   & \large \cgreen{A} \\
\normalfont19&\normalfont (3,2) & TCCG  & G \\
\normalfont20&\normalfont (1,2) &  TGAC  & C \\
\normalfont21&\normalfont (2,1) &  TG    & \large \cgreen{G} \\
\hline
\end{tabular}
\end{minipage}
}
\quad
\scalebox{0.78}{
\vtop{ \hbox{ \begin{tabular}{ |r|r|l|c||c||c||c||c|l|r| } 
\hline
&\multicolumn{3}{c||}{\normalfont dollar-EBWT}&\multicolumn{6}{c|}{\normalfont multidollar-EBWT}\\
\hline
\normalfont $i$&\normalfont GCA & \normalfont rotation & & \normalfont opt & \normalfont plus & \normalfont colex & \normalfont input & \normalfont rotation & \normalfont GSA   \\ 
\hline
\normalfont1&\normalfont (6,4) & \cgreen{\$}\cgray{CGA}   & \large \cgreen{A} & \large \cgreen{A} & \large \cgreen{A} & \large \cgreen{A} & \large \cgreen{A} & \cgreen{\$$_1$}\cgray{CTGA} &\normalfont (1,5) \\
\normalfont2&\normalfont(5,6) & \cgreen{\$}\cgray{CGACC} & \large \cgreen{C} & \large \cgreen{A} & \large \cgreen{A} & \large \cgreen{A} & \large \cgreen{G} & \cgreen{\$$_2$}\cgray{TG} &\normalfont (2,3) \\
\normalfont3&\normalfont(1,5) & \cgreen{\$}\cgray{CTGA}  & \large \cgreen{A} & \large \cgreen{A} & \large \cgreen{A} & \large \cgreen{A} & \large \cgreen{C} & \cgreen{\$$_3$}\cgray{GTCC} &\normalfont (3,5)  \\
\normalfont4&\normalfont(3,5) & \cgreen{\$}\cgray{GTCC}  & \large \cgreen{C} & \large \cgreen{G} & \large \cgreen{C} & \large \cgreen{C} & \large \cgreen{A} & \cgreen{\$$_4$}\cgray{TCA} &\normalfont (4,4) \\
\normalfont5&\normalfont(4,4) & \cgreen{\$}\cgray{TCA}   & \large \cgreen{A} & \large \cgreen{C} & \large \cgreen{C} & \large \cgreen{C} & \large \cgreen{C} & \cgreen{\$$_5$}\cgray{CGACC} &\normalfont (5,6) \\
\normalfont6&\normalfont(2,3) & \cgreen{\$}\cgray{TG}    & \large \cgreen{G} & \large \cgreen{C} & \large \cgreen{G} & \large \cgreen{G} & \large \cgreen{A} & \cgreen{\$$_6$}\cgray{CGA} &\normalfont (6,4) \\
\hline 
\normalfont7&\normalfont(6,3) & \red{A\$}\cgray{CG}   & \large \red{G} & \large \red{C} & \large \red{G} & \large \red{C} & \large \red{G} & \red{A\$$_1$}\cgray{CTG} &\normalfont (1,4) \\
\normalfont8&\normalfont(1,4) & \red{A\$}\cgray{CTG}  & \large \red{G} & \large \red{G} & \large \red{G} & \large \red{G} & \large \red{C} & \red{A\$$_4$}\cgray{TC} &\normalfont (4,3) \\
\normalfont9&\normalfont(4,3) & \red{A\$}\cgray{TC}   & \large \red{C} & \large \red{G} & \large \red{C} & \large \red{G} & \large \red{G} & \red{A\$$_6$}\cgray{CG} &\normalfont (6,3) \\
\hline 
\normalfont10&\normalfont(5,3) & ACC\$\cgray{CG} & G & G & G & G & G&ACC\$$_5$\cgray{CG} &\normalfont (5,3) \\
\hline 
\normalfont11&\normalfont(5,5) & \blue{C\$}\cgray{CGAC} & \large \blue{C} & \large \blue{C} & \large \blue{C} & \large \blue{C} & \large \blue{C} & \blue{C\$$_3$}\cgray{GTC} &\normalfont (3,4) \\
\normalfont12&\normalfont(3,4) & \blue{C\$}\cgray{GTC}  & \large \blue{C} & \large \blue{C} & \large \blue{C} & \large \blue{C} & \large \blue{C} & \blue{C\$$_5$}\cgray{CGAC} &\normalfont (5,5) \\
\hline 
\normalfont13&\normalfont(4,2) & CA\$\cgray{T}   & T & T & T & T & T &  CA\$$_4$\cgray{T} &\normalfont (4,2) \\
\hline 
\normalfont14&\normalfont(5,4) & \orange{CC\$}\cgray{CGA} & \large \orange{A} & \large \orange{T} & \large \orange{T} & \large \orange{A} & \large \orange{T}&\orange{CC\$$_3$}\cgray{GT} &\normalfont (3,3) \\
\normalfont15&\normalfont(3,3) & \orange{CC\$}\cgray{GT}  & \large \orange{T} &\large \orange{A} & \large \orange{A} & \large \orange{T} & \large \orange{A} &\orange{CC\$$_5$}\cgray{CGA} &\normalfont (5,4) \\
\hline 
\normalfont16&\normalfont(6,1) & CGA\$   & \$ &\$$_3$ &\$$_2$ &\$$_2$ & \$$_6$ & CGA\$$_6$ &\normalfont (6,1) \\
\hline 
\normalfont17&\normalfont(5,1) & CGACC\$ & \$  &\$$_6$ &\$$_5$&\$$_4$ &\$$_5$ & CGACC\$$_5$ &\normalfont (5,1) \\
\hline 
\normalfont18&\normalfont(1,1) & CTGA\$  & \$ &\$$_2$ &\$$_1$ &\$$_3$ &\$$_1$ & CTGA\$$_1$ &\normalfont (1,1) \\
\hline 
\normalfont19&\normalfont(2,2) & G\$\cgray{T}    & T & T &T& T & T&G\$$_2$\cgray{T} &  \normalfont (2,2) \\
\hline 
\normalfont20&\normalfont(6,2) & \violet{GA\$}\cgray{C}   & \large \violet{C}  & \large \violet{T} & \large \violet{T} &\large \violet{C}& \large \violet{T}&\violet{GA\$$_1$}\cgray{CT} &\normalfont (1,3) \\
\normalfont21&\normalfont(1,3) & \violet{GA\$}\cgray{CT}  & \large \violet{T} & \large \violet{C} & \large \violet{C} &\large \violet{T} &\large \violet{C}&\violet{GA\$$_6$}\cgray{C} &\normalfont (6,2) \\
\hline 
\normalfont22&\normalfont(5,2) & GACC\$\cgray{C} & C &C&C&C& C&GACC\$$_5$\cgray{C} &\normalfont (5,2) \\
\hline 
\normalfont23&\normalfont(3,1) & GTCC\$  & \$ &\$$_5$ &\$$_4$ &\$$_5$& \$$_3$ & GTCC\$$_3$ &\normalfont (3,1) \\
\hline 
\normalfont24&\normalfont(4,1) & TCA\$   & \$ &\$$_1$ &\$$_3$&\$$_1$& \$$_4$ & TCA\$$_4$ &\normalfont (4,1) \\
\hline 
\normalfont25&\normalfont(3,2) & TCC\$G  & G &G&G& G&G & TCC\$$_3$\cgray{G} &\normalfont (3,2) \\ 
\hline 
\normalfont26&\normalfont(2,1) & TG\$    & \$ &\$$_4$ &\$$_6$&\$$_6$& \$$_2$ & TG\$$_2$ &\normalfont (2,1) \\
\hline 
\normalfont27&\normalfont(1,2) & TGA\$\cgray{C}  & C &C&C&C& C&TGA\$$_1$\cgray{C} &\normalfont (1,2) \\
\hline
\end{tabular}}
}}

\vspace{2mm}
\caption{\EBWT-based BWT variants: from left to right, we show the extended BWT (\EBWT), the dollar-EBWT, and the multidollar-EBWT (using different string orderings) of the string collection ${\cal M} =  ({\texttt{CTGA},\texttt{TG},\texttt{GTCC},\texttt{TCA},\texttt{CGACC}, \texttt{CGA}})$. We use different colors to identify the characters preceding different identical suffixes. See Sections~\ref{sec:ebwt-based} and \ref{sec:opt} for more details. 
}
\label{tab:eBWT-based-running-example}
\end{table*}

\subsection{Concatenation-based transforms}\label{sec:concatenation-based}

In this section, we present the two concatenation-based transforms, \mdolBWT\ and \concat. See \cref{tab:concat-based-running-example} for an example.

\subparagraph{Multidollar-BWT} 
In this transformation, denoted $\mdolBWT$, given the string collection $\mathcal{M}=(T_1, \ldots, T_m)$, the strings are concatenated in the order in which they appear in the $m$-tuple, using $\$_1<\$_2<\ldots<\$_m$ as separators, and then the BWT is applied to the concatenated string. More formally, 
\[\mdolBWT(T_1,\ldots,T_m)=\BWT(T_1\$_1T_2\$_2\cdots T_m\$_m).\]

Unlike the \EBWT-based transforms, if $\pi_L$ is the standard permutation of the word $L=\dollarebwt({\cal M})$, 
then $\pi_L$ consists of a single cycle. 
Each string $T_d$ in the collection can be reconstructed, using the LF mapping, starting from the symbol $\$_d$
and terminating the reconstruction when the symbol $\$_{d-1}$ (for $d>1$) or $\$_m$ (for $d=1$) is reached. 

The order in which the strings appear in the collection can 
 significantly affect the output, giving rise to specific transforms. 
 See \cref{sec:opt} for more details. 

Note that sometimes the separators are added only implicitly when concatenating the strings, for example by maintaining a bitvector that marks the beginning of a new string. 
 Moreover, 
 some implementations of $\mdolBWT$ compute the $\BWT$ after appending an additional end-of-string symbol $\#$, which is smaller than all the characters of the alphabet and the dollar symbols, to the concatenation of the strings. 
 The resulting output differs from $\mdolBWT$ only in that the symbol $\$_m$ is added in the first position, and the symbol $\#$ replaces $\$_m$. The \mdolBWT($\cal M$) is constructed using the suffix array \SA\ of the concatenated string $T = T_1\$_1T_2\$_2\cdots T_m\$_m$ as follows: 
\begin{equation} \label{eq:mdolBWT}
\mdolBWT(\bigM)[i] = {\left\{ 
\begin{array}{ll} 
    T[\SA[i]-1] &  \text{ if } \SA[i]>1, \\ 
    \$_m       &  \text{ if } \SA[i]=1. 
\end{array}
\right. } 
\end{equation}

\begin{example}
Let us consider the collection ${\cal M} = ( \mathtt{CTGA},\mathtt{TG},\mathtt{GTCC},\mathtt{TCA},\mathtt{CGACC}, \mathtt{CGA})$. Then $\mdolBWT(\mathcal{M}) = 
\mathtt{AGCACAGCGGCCTTA\$_5\$_4\$_6TTCC\$_2\$_3G\$_1C}$, as shown in Table \ref{tab:concat-based-running-example}. 
The LF-mapping consists of a single cycle, which is 
$(1\ 7\ 20\ 27\ 18\ 6\ 9\ 21\ 16\ 5\ 12\ 15\ 10\ 22\ 17\ 4\ 8\ 13\ 24\ 3\ 11$
$\ 14\ 25\ 23\ 2\ 19\ 26)$. 
If ${\cal M''}$ contains the same strings in colexicographic order, i.e., $\mathcal{M''} $ $=(\mathtt{TCA},$ $\mathtt{CGA},$ $\mathtt{CTGA},$ $\mathtt{CGACC},$ $\mathtt{GTCC},$ $\mathtt{TG})$, 
then $\mdolBWT({\cal M''})=\mathtt{AAACCGCGGGCCTAT\$_1\$_3\$_2TCTC\$_4\$_6G\$_5C}$. 
If the end-of-string symbol $\#$ is appended to the concatenated strings, then the output of the transformation is $\mathtt{\$_6AGCACAGCGGCCTTA\$_5\$_4\#TTCC\$_2\$_3G\$_1C}$. 
\end{example}

\subparagraph{Concatenated-BWT}
This transformation, denoted \concat, concatenates all the strings into a single string using a single end-of-string symbol $\$$ as separator and appending one additional character, the final end-of-string symbol, here denoted by $\#$, which is smaller than all other characters. Then the classic BWT is applied to this string. 
This is the method implicitly applied when using a tool for classic BWT construction, so probably this is the most commonly used method. 
Given the collection $\mathcal{M}=(T_1, T_2, \ldots, T_m)$, 
\[\concat(T_1,\ldots,T_m)=\BWT(T_1\$T_2\$\cdots T_m\$\#).\]

The \concat\ can be computed using the \SA\ of the concatenation of the input strings, similarly to Eq.\ \eqref{eq:mdolBWT}, as follows: 
\begin{equation} \label{eq:concatBWT}
\concat(\bigM)[i] = {\left\{ 
\begin{array}{ll} 
    T[\SA[i]-1] &  \text{ if } \SA[i]>1, \\ 
    \#       &  \text{ if } \SA[i]=1. 
\end{array}
\right. } 
\end{equation}

Because of the extra end-of-string symbol, the additional suffix starting with \# is the smallest, and therefore, we haver an additional dollar in the first position of the transform. To facilitate the comparison with the other transforms, we denote by {\it adapted \concat}, the transform obtained by removing the first symbol from \concat\ and replacing the \# by \$.

\begin{example}\label{ex:2}
Let us consider the collection ${\cal M} = ( \mathtt{CTGA},\mathtt{TG},\mathtt{GTCC},\mathtt{TCA},\mathtt{CGACC}, \mathtt{CGA})$. Then $\concat(\mathcal{M})=
\mathtt{\$ACAGCAGCGGCCTAT\$\$\#TCTC\$\$G\$C}$, 
as shown in Table \ref{tab:concat-based-running-example}. 
The LF-mapping consists of a single cycle, which is 
$(1\ 2\ 8\ 21\ 17\ 3\ 12\ 15\ 11\ 23\ 18\ 4\ 9\ 14\ 25\ 6\ 13\ 16\ 26\ 24\ 5$
$\ 20\ 27\ 7\ 10\ 22\ 28\ 19)$. 
The adapted concatBWT is $\mathtt{ACAGCAGCGGCCTAT\$\$\$TCTC\$\$G\$C}$.
\end{example}

Note that the \#-symbol at the end of the string is a necessary addition, as without it, the LF-mapping, and thus, backward search, may not work correctly. This is because in that case, a suffix of the concatenated string may be a proper prefix of another suffix. This is the BWT variant treated in~\cite{CazauxR19,LiuZW16}. 

\begin{table*}[!h]
\centering
\ttfamily

\renewcommand{\arraystretch}{1.1}
\scalebox{0.8}{
\begin{tabular}{ |r|r|l|c||c|l|r|r| } 
\hline
&\multicolumn{3}{c||}{\normalfont multidollar-BWT}&\multicolumn{3}{c|}{\normalfont concatenated-BWT}&\\
\cline{2-7}
\normalfont $i$&\normalfont SA & \normalfont rotation & & & \normalfont rotation & \normalfont SA & \normalfont $i$   \\ 
\hline
& & & & \$ & \# & \normalfont28& \normalfont 1 \\
\hline
\normalfont1&\normalfont 5 & \cgreen{\$$_1$}\cgray{TG\$$_2$GTCC\$$_3$TCA\$$_4$CGACC\$$_5$CGA\$$_6$} & \large \cgreen{A} & \large \cgreen{A} & \cgreen{\$}\# & \normalfont27& \normalfont 2 \\
\normalfont2&\normalfont8 & \cgreen{\$$_2$}\cgray{GTCC\$$_3$TCA\$$_4$CGACC\$$_5$CGA\$$_6$} & \large\cgreen{G} & \large\cgreen{C} & \cgreen{\$}CGA\$\# & \normalfont23&\normalfont3 \\
\normalfont3&\normalfont13 & \cgreen{\$$_3$}\cgray{TCA\$$_4$CGACC\$$_5$CGA\$$_6$} & \large\cgreen{C} & \large\cgreen{A} & \cgreen{\$}CGACC\$CGA\$\# & \normalfont17&\normalfont4 \\
\normalfont4&\normalfont17 & \cgreen{\$$_4$}\cgray{CGACC\$$_5$CGA\$$_6$} & \large\cgreen{A} & \large\cgreen{G} & \cgreen{\$}GTCC\$TCA\$CGACC\$CGA\$\# & \normalfont8&\normalfont5 \\
\normalfont5&\normalfont23 & \cgreen{\$$_5$}\cgray{CGA\$$_6$} & \large\cgreen{C} & \large\cgreen{C} & \cgreen{\$}TCA\$CGACC\$CGA\$\# & \normalfont13 &\normalfont6\\
\normalfont6&\normalfont27 & \cgreen{\$$_6$} & \large\cgreen{A} & \large\cgreen{A} & \cgreen{\$}TG\$GTCC\$TCA\$CGACC\$CGA\$\# & \normalfont5 &\normalfont7\\
\hline 
\normalfont7&\normalfont4 & \red{A\$$_1$}\cgray{TG\$$_2$GTCC\$$_3$TCA\$$_4$CGACC\$$_5$CGA\$$_6$} & \large\red{G} & \large\red{G} & \red{A\$}\# & \normalfont26 &\normalfont8\\
\normalfont8&\normalfont16 & \red{A\$$_4$}\cgray{CGACC\$$_5$CGA\$$_6$} & \large\red{C} & \large\red{C} & \red{A\$}CGACC\$CGA\$\# &\normalfont 16 &\normalfont9\\
\normalfont9&\normalfont26 & \red{A\$$_6$} & \large\red{G} & \large\red{G} & \red{A\$}TG\$GTCC\$TCA\$CGACC\$CGA\$\# & \normalfont4 &\normalfont10\\
\hline 
\normalfont10&\normalfont20 & ACC\$$_5$\cgray{CGA\$$_6$} & G & G & ACC\$CGA\$\# & \normalfont20 &\normalfont11\\
\hline 
\normalfont11&\normalfont12 & \blue{C\$$_3$}\cgray{TCA\$$_4$CGACC\$$_5$CGA\$$_6$} & \large\blue{C} & \large\blue{C} & \blue{C\$}CGA\$\# &\normalfont 22 &\normalfont12\\
\normalfont12&\normalfont22 & \blue{C\$$_5$}\cgray{CGA\$$_6$} & \large\blue{C} & \large\blue{C} & \blue{C\$}TCA\$CGACC\$CGA\$\# &\normalfont 12 &\normalfont13\\
\hline 
\normalfont13&\normalfont15 & CA\$$_4$\cgray{CGACC\$$_5$CGA\$$_6$} & T & T & CA\$CGACC\$CGA\$\# &\normalfont 15 &\normalfont14\\
\hline 
\normalfont14&\normalfont11 & \orange{CC\$$_3$}\cgray{TCA\$$_4$CGACC\$$_5$CGA\$$_6$} & \large\orange{T} & \large\orange{A} & \orange{CC\$}CGA\$\# &\normalfont 21 &\normalfont15\\
\normalfont15&\normalfont21 & \orange{CC\$$_5$}\cgray{CGA\$$_6$} & \large\orange{A} & \large\orange{T} & \orange{CC\$}TCA\$CGACC\$CGA\$\# &\normalfont 11 &\normalfont16\\
\hline 
\normalfont16&\normalfont24 & CGA\$$_6$ & \$$_5$ & \$ & CGA\$\# &\normalfont 24 & \normalfont17\\
\hline
\normalfont17&\normalfont18 & CGACC\$$_5$\cgray{CGA\$$_6$} & \$$_4$ & \$ & CGACC\$CGA\$\# &\normalfont 18 & \normalfont18\\
\hline
\normalfont18&\normalfont1 & CTGA\$$_1$\cgray{TG\$$_2$GTCC\$$_3$TCA\$$_4$CGACC\$$_5$CGA\$$_6$} & \$$_6$  & \# & CTGA\$TG\$GTCC\$TCA\$CGACC\$CGA\$\# &\normalfont 1 & \normalfont19\\
\hline
\normalfont19&\normalfont7 & G\$$_2$\cgray{GTCC\$$_3$TCA\$$_4$CGACC\$$_5$CGA\$$_6$} & T & T & G\$GTCC\$TCA\$CGACC\$CGA\$\# &\normalfont 7 & \normalfont20\\
\hline 
\normalfont20&\normalfont3 & \violet{GA\$$_1$}\cgray{TG\$$_2$GTCC\$$_3$TCA\$$_4$CGACC\$$_5$CGA\$$_6$} & \large\violet{T} & \large\violet{C} & \violet{GA\$}\# &\normalfont 25 & \normalfont21\\
\normalfont21&\normalfont25 & \violet{GA\$$_6$} & \large\violet{C} & \large\violet{T} & \violet{GA\$}TG\$GTCC\$TCA\$CGACC\$CGA\$\# &\normalfont 3 & \normalfont22\\
\hline 
\normalfont22&\normalfont19 & GACC\$$_5$\cgray{CGA\$$_6$} & C & C & GACC\$CGA\$\# & \normalfont19 & \normalfont23\\
\hline
\normalfont23&\normalfont9 & GTCC\$$_3$\cgray{TCA\$$_4$CGACC\$$_5$CGA\$$_6$} & \$$_2$ & \$ & GTCC\$TCA\$CGACC\$CGA\$\# &\normalfont 9 & \normalfont24\\
\hline
\normalfont24&\normalfont14 & TCA\$$_4$\cgray{CGACC\$$_5$CGA\$$_6$} & \$$_3$ & \$ & TCA\$CGACC\$CGA\$\# & \normalfont14 & \normalfont25\\
\hline
\normalfont25&\normalfont10 & TCC\$$_3$\cgray{TCA\$$_4$CGACC\$$_5$CGA\$$_6$} & G & G & TCC\$TCA\$CGACC\$CGA\$\# &\normalfont 10 & \normalfont26\\
\hline
\normalfont26&\normalfont6 & TG\$$_2$\cgray{GTCC\$$_3$TCA\$$_4$CGACC\$$_5$CGA\$$_6$} & \$$_1$ & \$ & TG\$GTCC\$TCA\$CGACC\$CGA\$\# &\normalfont 6 &\normalfont 27\\
\hline
\normalfont27&\normalfont2 & TGA\$$_1$\cgray{TG\$$_2$GTCC\$$_3$TCA\$$_4$CGACC\$$_5$CGA\$$_6$} & C & C & TGA\$TG\$GTCC\$TCA\$CGACC\$CGA\$\# &\normalfont 2 & \normalfont 28\\
\hline
\end{tabular}
}

\vspace{2mm}
\caption{Concatenation-based BWT variants: the multidollar-BWT (left) and the concat-BWT (right) of the string collection ${\cal M} = ( {\texttt{CTGA}, \texttt{TG}, \texttt{GTCC}, \texttt{TCA}, \texttt{CGACC}, \texttt{CGA}})$. We use different colors to identify the characters preceding different identical suffixes. See \cref{sec:concatenation-based} for more details.}
\label{tab:concat-based-running-example}
\end{table*}

\subsection{A short literature overview of BWT variants}

Until a few years ago, all methods and algorithms extending the Burrows-Wheeler Transform (BWT) to string collections were considered equivalent. In the previous subsections, we described different types of approaches and the specific characteristics of each BWT variant. In this subsection, we survey the main papers where the BWT variants have been treated, often using different algorithmic strategies. The corresponding tools are listed in Table 1.

Linear-time algorithms for the computation of the 
original EBWT variant were recently presented in \cite{BCLRS_SPIRE2021_ebwt,Olbrich_lfgsaca,BannaiKKP24}. In particular, in \cite{BCLRS_SPIRE2021_ebwt}, an adaptation of the well-known Suffix Array Induced Sorting (SAIS) algorithm of Nong et al. \cite{Nong2011} is used. In \cite{BannaiKKP24}, the computation is done via computing the Bijective BWT (BBWT) of a string which is obtained from the input multiset. In \cite{Olbrich_lfgsaca}, a non-recursive suffix array construction is used.

The \dollarebwt\ variant is studied in \cite{DominguezN21,HoltMcMillan2014}. In \cite{DominguezN21}, the computation is based on the grammar compression of the text. In \cite{HoltMcMillan2014}, a merging strategy is used to compute the BWT of a collection by successively merging the BWTs of smaller subsets.
The $\multiEBWT$ and \mdolBWT\ variants, although based on different constructions---via the EBWT and via concatenation, respectively---produce essentially the same output string when the indices associated with the dollar symbols are ignored. Indeed, their output strings coincide up to the renaming of the dollar symbols, as detailed in Section \ref{sec:properties}. Notably, the LF-mapping associated with $\multiEBWT$ decomposes into a number of cycles equal to the number of input strings, whereas for \mdolBWT\ it forms a single cycle. However, the two transforms are easily related: one can be obtained from the other by shifting by one the indices associated with the dollar symbols, as stated in Lemma \ref{lemma:mdollar}. Therefore, any tool computing one of these two BWT variants can be easily adapted to compute the other. For this reason, in Table 1 where the indices associated to dollar symbols are ignored, we grouped together all tools that compute $\multiEBWT$ and/or \mdolBWT. 

More in detail, the computation of $\multiEBWT$ is explicitly treated in \cite{BauerCoxRosoneTCS2013,Li14a,Siren16,Louza2017d,EgidiLMT19_egap,BonizzoniVPPR19,LTGPR2020gsufsort,BonizzoniDVPPR_2021,DiazNavarro2023_grlBWT,ropebwt3}. Different algorithmic strategies are described in these papers: some algorithms implicitly sort the suffixes or build the generalized suffix array incrementally; others compute the transform by merging partial BWTs obtained from subsets of the collection; and some---such as the approach described in \cite{LTGPR2020gsufsort}---adopt a hybrid strategy, constructing the GSA starting from a single concatenated string and then adapting it to the collection setting.
Instead, the  \mdolBWT\ variant is explicitly used in \cite{bentley_et_alESA2020opt} to address the problem of finding the optimal order of concatenation of the strings, aiming at minimizing the number of runs in the resulting BWT.

The \concat\ variant, which uses a single dollar symbol as a separator and appends a final hash character, is adopted in \cite{BGKLMM2019_bigBWT,OlivaGB23,Masillo23}.
This is also the variant implicitly computed by tools originally designed for single-string BWT construction.

\section{Combinatorial properties of the different BWT variants}\label{sec:properties}

In this section, we study combinatorial properties of the five BWT variants introduced in \cref{sec:BWTvariants}. We give an overview in Table~\ref{tab:overview}. Full proofs for Sections~\ref{sec:separators} and~\ref{sec:interesting} can be found in the Supplementary Material of~\cite{CenzatoLiptak_bioinformatics2024}. 
Section~\ref{sec:dynamicity} is new, and Section~\ref{sec:output_order} is a generalization of the respective results in~\cite{CenzatoL22}. 

First notice that all BWT variants except the \EBWT append an end-of-string symbol (resp.\ separator-symbol) at the end of each input string. For this reason, we will in the following refer to these BWT variants as {\em separator-based} BWT variants. 

\begin{table*}[htbp]
\begin{center}
\scalebox{0.83}{
\begin{tabular}{l|p{5.9cm}|l|c|c}
\BWT\ variant & example & order of shared  & input order & dynami-\\
& & suffixes & dependence & city \\ 
\hline \hline
\multicolumn{4}{l}{\em \hspace{-4mm} \EBWT-based}\\
\hline
    $\eBWT({\cal M})$ & {\tt G\red{GGC}T\orange{A}\blue{C}\orange{T}\green{CA}\blue{C}\green{A}C\violet{CT}\green{C}T\green{A}GC\green{G} } & omega-order & no & yes \\
    $\dollarebwt({\cal M})$ & {\tt \green{ACACAG}\red{GGC}G\blue{CC}T\orange{AT}\$\$\$T\violet{CT}C\$\$G\$C} & lexicographic order & no  & yes  \\
    $\mdolEBWT({\cal M})$ & {\tt \green{AGCACA}\red{GCG}G\blue{CC}T\orange{TA}\$$_6$\$$_5$\$$_1$T\violet{TC}C\$$_3$\$$_4$G\$$_2$C} & input order & yes & yes\\
    \hline
    \multicolumn{4}{l}{\em \hspace{-4mm} concatenation-based}\\
    \hline
    $\mdollar({\cal M})$ & {\tt \green{AGCACA}\red{GCG}G\blue{CC}T\orange{TA}\$$_5$\$$_4$\$$_6$T\violet{TC}C\$$_2$\$$_3$G\$$_1$C } & input order & yes & yes\\
    $\concat({\cal M})$ & {\tt \$\green{ACAGCA}\red{GCG}G\blue{CC}T\orange{AT}\$\$\#T\violet{CT}C\$\$G\$C} & 
    lex.\ order of subseq.& \\
    &&  \; string in input & yes  & no\\
 \hline 
\end{tabular}
}
\end{center}
\vspace{2mm}
\caption{
Overview of some properties of the BWT\ variants considered in this paper on the string collection ${\cal M} = ( {\tt \texttt{CTGA},\texttt{TG},\texttt{GTCC},\texttt{TCA},\texttt{CGACC}, \texttt{CGA} })$. The colors in the example BWTs correspond to SAP intervals in separator-based variants, while the same characters are highlighted in the \EBWT\ for showing their positions, see \cref{sec:interesting}. Dynamicity is w.r.t.\ removal and addition of strings to the collection, see \cref{sec:dynamicity}. The order of shared suffixes is discussed in \cref{sec:output_order}.
\label{tab:overview}}
\end{table*}

\subsection{Differences between $\eBWT$ and separator-based variants}\label{sec:separators}

In all separator-based BWT variants, the $m$-length prefix of the resulting transform consists of a permutation of the last characters of the input strings $T_d$, since the smallest rotations are those starting with the separators. In the \eBWT, on the other hand, these $m$ characters appear interspersed with the other characters (see Table~\ref{tab:overview}). In general, adding separators introduces a distinction, not present in the \eBWT, between occurrences of a substring as suffix and otherwise: in the \eBWT, the left-context of a substring appears according to the omega-order of the corresponding rotations, while in the separator-based variants, the occurrences that are suffixes appear in one block at the start. 

To demonstrate this, consider the substring {\tt CC} in our running example. As can be seen in Table~\ref{tab:eBWT-based-running-example}, {\tt CC} occurs twice as suffix and once not as a suffix (in this case, this is a circular occurrence). The two preceding characters of the suffix occurrences, {\tt A} and {\tt T}, form one block in all separator-based BWTs, while in the \eBWT, they are intermixed with the preceding {\tt C} of the non-suffix occurrence. A similar observation can be made about the two occurrences of {\tt C} as suffix, whose preceding characters (both {\tt C}) appear in once block in the separator-based BWT variants, while they are intermixed with all other occurrences of {\tt C} in the \eBWT. 

Moreover, observe that adding separator-symbols changes the order of the strings. In particular, appending \$'s to the strings has the consequence that no string is a proper prefix of another, and therefore, the omega-order 
and the lexicographic order on these strings coincide. This is true whether the same separator-symbol or different symbols are used. 
An example is given in Figure~\ref{fig:ebwt-variants}, where we also show the difference between $\mdolEBWT({\cal M})$ and $\mdolBWT({\cal M})$. 

\begin{figure}[htpb]
\centering
\begin{tabular}{l@{\hspace{.1cm}}l@{\hspace{0.6cm}}|l@{\hspace{.1cm}}l@{\hspace{0.6cm}}|l@{\hspace{.1cm}}l@{\hspace{0.6cm}}|l@{\hspace{.1cm}}l@{\hspace{0.1cm}}}

\multicolumn{2}{l|}{\eBWT($\{{\tt GTC,GT}\}$)} & 
\multicolumn{2}{l|}{\eBWT($\{${\tt GTC\$,GT\$}$\}$)} & 
\multicolumn{2}{l|}{\eBWT($\{${\tt GTC\$$_1$,GT\$$_2$}$\}$)} &  
\multicolumn{2}{l}{$\BWT({\tt GTC}${\tt \$}$_1{\tt GT}${\tt \$}$_2)$} \\
\hline
{\tt CGT} \phantom{blablabla} & {\tt T} & {\tt \$GT} \phantom{blablabla} & {\tt T} & {\tt \$$_1$} \phantom{blablabla} & {\tt C} & {\tt \$$_1$GT\$$_2$GTC} \phantom{bla} & {\tt C}\\
{\tt GTC} & {\tt C} & {\tt \$GTC} & {\tt C} & {\tt \$$_2$} & {\tt T} & {\tt \$$_2$GTC\$$_1$GT} & {\tt T}\\
{\tt GT} & {\tt T} & {\tt C\$GT} & {\tt T} & {\tt C\$$_1$} & {\tt T} & {\tt C\$$_1$GT\$$_2$GT} & {\tt T}\\
{\tt TCG} & {\tt G} & {\tt GT\$} & {\tt \$} & {\tt GT\$$_2$} & {\tt \$$_2$} & {\tt GT\$$_2$GTC\$$_1$} & {\tt \$$_1$}\\
{\tt TG} & {\tt G} & {\tt GTC\$} & {\tt \$} & {\tt GTC\$$_1$} & {\tt \$$_1$} & {\tt GTC\$$_1$GT\$$_2$} & {\tt \$$_2$}\\
&& {\tt T\$G} & {\tt G} & {\tt T\$$_2$} & {\tt G} & {\tt T\$$_2$GTC\$$_1$G} & {\tt G}\\
&& {\tt TC\$G} & {\tt G} & {\tt TC\$$_1$} & {\tt G} & {\tt TC\$$_1$GT\$$_2$G} & {\tt G}\\
\hline 
\eBWT(${\cal M}$) & & \dollarebwt(${\cal M}$) & & \mdolEBWT(${\cal M}$) && \mdolBWT(${\cal M}$) \\ 
\end{tabular}
\caption{The \EBWT of the three sets $\{{\tt GTC,GT}\}$, $\{${\tt GTC\$,GT\$}$\}$, and $\{${\tt GTC\$$_1$,GT\$$_2$}$\}$, and the BWT of the concatenated string ${\tt GTC\$_1GT\$_2}$, corresponding to \eBWT, \dollarebwt, \mdolEBWT, and \mdolBWT\ of the same input. As can be seen, the result is different in each case. See Section~\ref{sec:separators} for a discussion.
}
\label{fig:ebwt-variants}
\end{figure}

Let us denote by $\same$ that two transforms are equal up to renaming the dollar symbols. 
The following lemma
states that the transformation $\mdolBWT$ and $\multiEBWT$ produce the same string, up to renaming dollars, and that $\dollarebwt$ is equivalent to first lexicographically sorting the input strings, and then applying either the $\mdolBWT$ or the $\mdolEBWT$.

\begin{lemma}\label{lemma:mdollar}
Let ${\cal M} = (T_1,T_2,\ldots,T_m)$ be a string collection. Then 

\begin{enumerate}
\item $\mdollar({\cal M}) \same \mdollarE({\cal M})$. Moreover, \$$_d$ in $\mdollarE({\cal M})$ is replaced by \$$_{d-1}$ in $\mdollar({\cal M})$, and \$$_1$ by \$$_m$. 
Formally, $\mdollar({\cal M}) = \eBWT(\{ T_d\$_{d-1}\mid d=1,\ldots,m, \text{ with } T_0 = T_m\})$. 
\item $\dollarebwt({\cal M}) \same \mdollar(\lex({\cal M})) \same \mdollarE(\lex({\cal M}))$, where $\lex({\cal M})$ denotes the lexicographic order of the strings in ${\cal M}$. 
\end{enumerate}
\end{lemma}

In the following section, we will concentrate on separator-based BWT variants only and show how they differ from each other. 

\subsection{Interesting intervals and SAP-intervals}\label{sec:interesting}

Let us call a string $U$ a {\em shared suffix} if there exist at least two strings $T_d,T_{d'}\in{\cal M}$, $d\neq d'$,
such that $U$ is a suffix of both $T_d$ and $T_{d'}$. Let $b$ denote the lexicographic rank of the smallest rotation 
beginning with $U\$$ (where \$ now stands for {\em any} separator-symbol) and $e$ that of the largest. In other words,  $[b,e]$ is the SA-interval of $U\$$. We refer to $[b,e]$ as an {\em SAP-interval}, 
following~\cite{CoxBJR_2012_sap}\footnote{SAP stands for ``same as previous''}. If there exist $d\neq d'$ such that $U$ is a suffix of both $T_i$ and $T_j$ and the preceding character in $T_d$ is distinct from the preceding character in $T_{d'}$, then $[b,e]$ is called an {\em interesting interval}~\cite{CenzatoL22,CenzatoLiptak_bioinformatics2024}. 

The next lemma follows from the fact that no two distinct substrings ending in $\$$ can be one the prefix of the other. 

\begin{lemma}
Any two distinct interesting intervals are disjoint. Moreover, any two distinct SAP-intervals are disjoint. 
\end{lemma}

Interesting intervals allow us to locate precisely how and where two separator-based BWT variants of the same multiset can differ, which we state formally in the next proposition. For the proof, see the Supplemental Material of~\cite{CenzatoLiptak_bioinformatics2024}. 

\begin{proposition}\label{prop:int_intervals}
Let $L_1$ and $L_2$ be two separator-based BWTs of the same multiset ${\cal M}$. 

\begin{enumerate} 
\item If $L_1[i] \neq L_2[i]$ then  $i \in [b,e]$ for some interesting interval $[b,e]$. 
\item \label{prop2} Let ${\cal I}_1$ resp.\ ${\cal I}_2$ be the sets of the positions of the dollars in $L_1$ resp.\ $L_2$. If ${\cal I}_1\neq {\cal I}_2$ then there exist $d\neq d'$ such that $T_d$ is a proper suffix of $T_{d'}$.  
\end{enumerate}
\end{proposition}

Thus, the differences between two separator-based BWT variants of the same multiset can be found only in interesting intervals (Property 1). To see Property 2, consider the following example: 

\begin{example}
Consider the \multiEBWT\ of the same multiset $\{{\tt GCA, CA}\}$, given in two different orders: $\multiEBWT({\tt GCA, CA}) = \mathtt{AACCG\$_2\$_1}$, while 
$\multiEBWT({\tt CA, GCA}) = \mathtt{AACC\$_1G\$_2}$. This is because the interesting interval $[5,6]$ associated with the shared suffix $\mathtt{CA}$ contains a dollar and a {\tt G}, which change place according to the input order. 
\end{example} 

\cref{prop:int_intervals} implies that the differences between two separator-based BWT variants of the same multiset can be fully explained based on what rule is used to break ties for shared suffixes. We will see in \cref{sec:output_order} how the different BWT variants determine this tie-breaking rule.

\subsection{Dynamicity of BWT variants}\label{sec:dynamicity}

In this subsection, we study the dynamicity of the five BWT variants presented: what happens to the transforms if one or more of the strings are removed from (or, symmetrically, added to) the collection? Are the transforms robust to these operations? In other words, is the relative order of characters preserved when strings are added or deleted? These characteristics are particularly important for adaptive indexing structures, as they could allow efficient updates when the input changes dynamically. 

We will see that all except the \concat\ are robust w.r.t.\ removal or addition of strings. 

We recall a string $S$ is a \emph{subsequence} of a string $T$ if $S$ can be obtained from $T$ by deleting zero or more characters without  changing the order of the remaining characters. More formally, given the string $T=T[1..|T|]$, a string $S=S[1..|S|]$ is a subsequence of $T$, denoted $S\sqsubseteq T$, if there exists a sequence of integers $i_1, i_2, \ldots, i_{|S|}$, such that $1\leq i_1< i_2 < \ldots < i_{|S|}\leq |T|$ and $S[j]=T[i_j]$ for every $j=1,\ldots,|S|$.

The following lemma follows from the fact that, for $T_d\in {\cal M}$, any two cyclic rotations (resp.\ suffixes) of  string $T_d$ preserve their relative lexicographic order when they are considered in the $\omega$-sorted (resp.\ lexicographically sorted) list of all cyclic rotations (resp. suffixes) of the string collection ${\cal M}$.

\begin{lemma}
\label{le:bwt_subsequence}
Given a multiset $\mathcal{M}$, for every string $T_d \in \mathcal{M}$, 
\begin{enumerate}
\item $\bwt(T_d)\sqsubseteq \eBWT(\mathcal{M})$, 
\item $\bwt(T_d\$)\sqsubseteq \dollarebwt(\mathcal{M})$, 
\item $\bwt(T_d\$)\sqsubseteq\mdolEBWT(\mathcal{M})$, up to renaming the character $\$$,
\item $\bwt(T_d\$)\sqsubseteq\mdolBWT(\mathcal{M})$, up to renaming the character $\$$,
\item $\bwt(T_d\$)\sqsubseteq\concat(\mathcal{M})$, up to renaming the character $\$$.
\end{enumerate}
\end{lemma}

When subsets of the string collection are considered, rather than a single string, the following proposition shows that Lemma \ref{le:bwt_subsequence} can be generalized for $\EBWT$ and $\dollarebwt$.

\begin{proposition}\label{prop:dyn1}
Let $\mathcal{M}$ be a multiset of strings and let $\mathcal{S}\subseteq \mathcal{M}$. Then: 
\begin{enumerate}
\item $\eBWT(\mathcal{S})\sqsubseteq \eBWT(\mathcal{M})$, and 
\item $\dollarebwt(\mathcal{S})\sqsubseteq \dollarebwt(\mathcal{M})$. 
\end{enumerate}
\end{proposition}

In order to study the behavior of the input order dependent BWT variants when sub-collections of strings are considered, we use the following definition. Given an ordered collection $\mathcal{M}=(T_1, T_2, \ldots, T_m)$, the ordered collection $\mathcal{S}=(S_1, S_2, \ldots, S_p)$ is a \emph{sub-collection} of $\mathcal{M}$ (denoted $\mathcal{S}\preceq \mathcal{M}$) if there exists  a sequence of integers $i_1, i_2, \ldots, i_{p}$, such that $1\leq i_1 < i_2 < \ldots < i_{p}\leq m$ and $S_j=T_{i_j}$ for every $j=1,\ldots,p$.

In the following proposition, we consider $\mdolEBWT$ and $\mdolBWT$. In both of these transformations, the relative order of two identical suffixes is determined by the order of the distinct dollar symbols. Therefore, if the sub-collection of strings maintains the same relative order, then the output of each transformation applied to the sub-collection is a subsequence of the output obtained with the entire string collection.

\begin{proposition}\label{prop:dyn2}
Let $\mathcal{M}$ be an ordered string collection and let $\mathcal{S}\preceq \mathcal{M}$. Then:
\begin{enumerate}
\item $\mdolEBWT(\mathcal{S})\sqsubseteq \mdolEBWT(\mathcal{M})$, up to renaming the dollar symbols used for the strings in $\mathcal{S}$, and 
\item $\mdolBWT(\mathcal{S})\sqsubseteq \mdolBWT(\mathcal{M})$, up to renaming the dollar symbols used for the strings in $\mathcal{S}$. 
\end{enumerate}

\end{proposition}

The previous two propositions show that all the considered BWT-variants can be used as basis of dynamic compressed indexing structures, since they allow the removal of a string from the collection simply by deleting the characters of the string from the output of the transformation on the full string collection. Similarly, they enable the use of the merge of the transforms on sub-collections to obtain the output of a collection containing all the strings.

The following example shows that, on the other hand, the BWT-variant we refer to as \concat, i.e., concatenating the input strings using the same dollar-symbol to separate them, does not satisfy this property.

\begin{example}
Let us consider the collection $\mathcal{M}=(\mathtt{CCA},\mathtt{ACA}, \mathtt{TCA})$. Then $\concat(\mathcal{M})=\mathtt{\$AAACCC\$TCA\#\$}$. When the sub-collection $\mathcal{S}=(\mathtt{CCA},\mathtt{ACA})$ is considered, then $\concat(\mathcal{S})$ $=\mathtt{\$AACC\$AC\#}$, which is not a subsequence of $\concat(\mathcal{M})$.
\end{example}

\subsection{Input order and output order}\label{sec:output_order}

It is clear that the first $m$ characters of a separator-based BWT variant constitute an SAP-interval, namely the one corresponding to $U=\epsilon$. Since this $m$-length prefix contains every string $T_d$ exactly once, it determines a permutation of the input strings. This is the permutation of the indices $1\leq d \leq m$ found in the $m$ first entries of $\GCA$, i.e., in $\GCA[1..m]$.\footnote{Or, equivalently, in the document array DA.} 
This permutation then determines the order in which the respective suffixes appear in the interesting intervals, which, by~\cref{prop:int_intervals}, determines the final transform.

As we have a multiset in input, the relative order of equal strings $T_d=T_{d'}$, where $d\neq d'$, does not matter. We can thus use a meta-string to model the input and output permutations of ${\cal M}$, as follows. Let $m'\leq m$ denote the number of distinct strings in ${\cal M}$. We define a meta-string $t$ of length $m$, with characters from an ordered alphabet of size $m'$, to model the {\em input permutation} of ${\cal M} = (T_1,T_2,\ldots,T_m)$, by replacing every string $T_d$ by a meta-character according to its lexicographic rank among all distinct input strings. For example, for ${\cal M} = ({\tt ACA,TGA,ACA,GAA,TGA,TGA})$, we get $t = {\tt acabcc}$. Now  the $m$-length prefix of the BWT, corresponding to the {\em output permutation}, can be seen as another meta-string, which is, in fact, a permutation of $t$. As this string depends on the BWT variant employed, each of these variants can be viewed as a string transform, acting on the input meta-string $t$. We will denote these transforms as $\pi_{dolE}, \pi_{mdolE}, \pi_{mdol}$, and $\pi_{concat}$, respectively. 

\begin{example}\label{ex:order}
    Let ${\cal M} = ({\tt ACA,TGA,ACA,GAA,TGA,TGA})$. Then $t = {\tt acabcc}$ and $\pi_{mdolE}(t) = \pi_{mdol}(t) = {\tt acabcc}$, $\pi_{dolE}(t) = {\tt aabccc}$, and $\pi_{conc}(t) = {\tt ccacab}$. 
\end{example}

\begin{figure}[htpb]
\centering
\scalebox{0.96}{
\begin{tabular}{lc|lc|lc|lc}

\multicolumn{1}{l}{\mdolEBWT} & \multicolumn{1}{l|}{$\pi_{mdolE}$} &
\multicolumn{1}{l}{\mdolBWT} & \multicolumn{1}{l|}{$\pi_{mdol}$} &
\multicolumn{1}{l}{\dollarebwt} & \multicolumn{1}{l|}{$\pi_{dolE}$}  &
\multicolumn{1}{l}{\concat} & \multicolumn{1}{l}{$\pi_{conc}$} \\
\hline
{\tt \$$_1$ACA}  & {\tt a} & {\tt \$$_1$TGA\cgray{\$$_2$ACA..}} & {\tt a} & {\tt \$ACA} & {\tt a} & {\tt \$\#} & {\tt c}\\

{\tt \$$_2$TGA} & {\tt c} & {\tt \$$_2$ACA\cgray{\$$_3$GAA..}} & {\tt c} & {\tt \$ACA} & {\tt a} & {\tt \$ACA\$GAA\$..} & {\tt c}\\

{\tt \$$_3$ACA} & {\tt a} & {\tt \$$_3$GAA\cgray{\$$_4$TGA..}} & {\tt a} & {\tt \$GAA} & {\tt b} & {\tt \$GAA\$TGA\$..} & {\tt a}\\

{\tt \$$_4$GAA} & {\tt b} & {\tt \$$_4$TGA\cgray{\$$_5$TGA..}} & {\tt b} & {\tt \$TGA} & {\tt c} & {\tt \$TGA\$\# } & {\tt c}\\

{\tt \$$_5$TGA} & {\tt c} & {\tt \$$_5$TGA\cgray{\$$_6$}} & {\tt c} & {\tt \$TGA} & {\tt c} & {\tt \$TGA\$ACA\$..} & {\tt a}\\

{\tt \$$_6$TGA} & {\tt c} & {\tt \$$_6$} & {\tt c} & {\tt \$TGA} & {\tt c} & {\tt \$TGA\$TGA\$\#} & {\tt b}\\

\hline 
\end{tabular}
}
\caption{The first $m$ rotations or suffixes and the output permutations for all separator-based $\BWT$s of ${\cal M} = (\tt ACA,TGA,ACA,GAA,TGA,TGA)$. }
\label{fig:ebwt-variants-2}
\end{figure}

It is easy to see that both $\pi_{mdolE}(t)$ and $\pi_{mdol}(t)$ are equal to $t$, since the dollar-symbols are ordered according to the input order. For the \dollarebwt, the rank of {\tt \$}$T_d$ equals the lexicographic rank of $T_d$ among all input strings (Lemma~\ref{lemma:mdollar}), i.e., $\pi_{dolE}(t)$ is the lexicographically sorted permutation of the input string $t$. The situation is more complex for the \concat. Since {\tt \#} is the smallest character and it is in the final position of the concatenated string $T_1${\tt \$}$T_2${\tt \$}$\cdots T_m${\tt \$\#}, the last string $T_m$ of the input will be the first, while for the other strings $T_d$, the lexicographic rank {\em of the following string} $T_{d+1}$ decides the order. Moreover, if $T_d$ and $T_{d'}$ are followed by two equal strings $T_{d+1}=T_{d'+1}$, then the lexicographic order of the strings following the following strings will break the tie, and so on. In other words, $\pi_{concat}$ is essentially the BWT of $t$. 
(We thank Massimiliano Rossi for this insight.) 

\begin{lemma}\label{lemma:concat}
Let $t$ be the input order meta-string of ${\cal M} = (T_1,\ldots, T_m)$. Then $\pi_{concat}(t) = \BWT^*(t)$, where for a string $u$,  $\BWT^*(u)$ is defined as $\BWT(u${\tt \$}$)$ from which the end-of-string symbol {\tt \$} has been removed.
\end{lemma}

\begin{example}\label{ex:concat}
    Continuing \cref{ex:order}, $\BWT(t{\tt \$}) = \BWT({\tt acabcc\$})$ $ = {\tt cc\$acab},$ and therefore, $\BWT^*(t) = {\tt ccacab}. $
\end{example}

Lemma~\ref{lemma:concat} has important consequences for the BWT variant \concat. First, it means that in most cases, the transforms \mdolBWT\ and \concat\ will produce different outputs, given the same input. This is because $\pi_{concat}(t)[1] = t[m] = \pi_{mdol}(t)[m]$, and therefore, 
if the strings in ${\cal M}$ are all distinct, then for any input order, the output permutations induced by \mdollar\ and \concat\ are distinct: in this case, $t[1]\neq t[m]$, and thus $\pi_{mdol}(t) \neq \pi_{concat}(t)$ holds for all $t$. This means that, in whatever order the strings are given, on most string sets the resulting transforms \mdollar\ and \concat\ will differ. 

Another consequence is that \concat\ cannot produce all possible transforms. Already for $m=3$, on an input collection of three distinct strings, the  mapping corresponding to \concat\ is not surjective, since 
the mapping $\pi_{concat}$ for $m'=m=3$ is as follows: ${\tt abc} \mapsto {\tt cab}$, ${\tt acb} \mapsto {\tt bca}$, ${\tt cab} \mapsto {\tt bca}$, ${\tt bac} \mapsto {\tt cba}$, ${\tt bca} \mapsto {\tt acb}$, and ${\tt cba} \mapsto {\tt abc}$. In particular, no $t$ maps to ${\tt bac}$. 

It has been shown experimentally that more than half of binary and ternary strings of length between $10$ and $20$ do not lie in the image of the function $\BWT^*$, with the percentage of those not in the image increasing with increasing length~\cite{GiulianiILPST21}.  These results seem to indicate that, in fact, the majority of multi-permutations cannot be produced by \concat\ (for a concrete example, see \cref{sec:opt}).

On the other hand, it follows from Lemmas~\ref{lemma:mdollar} and~\ref{lemma:concat} that every separator-based transform can be simulated by both \mdolBWT\ and \mdolEBWT, in the sense that there exists an input order of ${\cal M}$ such that the resulting \mdollar\ and \mdollarE\ will be equal to the transform, up to possibly renaming dollars. We summarize: 

\begin{theorem}\label{thm:multi_simulate}
Given a string collection ${\cal M} = (T_1,T_2,\ldots,T_m)$ and $L$ the output of a separator-based BWT variant on ${\cal M}$,  
there exists a permutation $\tau$ of ${\cal M}$ such that 
\[L \same \mdollarE(T_{\tau(1)},T_{\tau(2)}, \ldots, T_{\tau(m)})\same \mdollar(T_{\tau(1)},T_{\tau(2)}, \ldots, T_{\tau(m)}).\]
\end{theorem}

We will use the insight of \cref{thm:multi_simulate} in the next section.

\section{Reducing the number of runs}\label{sec:opt}

Recall that $L\same L'$ if the $L$ and $L'$ are equal up to renaming the dollar symbols. In \cref{sec:properties}, we showed that all separator-based BWT variants can be simulated by \mdolEBWT\ or by \mdolBWT\ (Thm.~\ref{thm:multi_simulate}): if $L$ is the output of one of these variants, then the input collection can be permuted in such a way that $\multiEBWT$ and $\mdolBWT$ will produce a transform which differs from $L$ only in the denomination of the dollar signs.  This justifies the following definition. 

Given a string collection ${\cal M}$, let $\mathfrak{S}_{\cal M}$ denote the 
string family obtained by applying $\multiEBWT$ to all possible orders of the strings in ${\cal M}$. Formally, for ${\cal M} = (T_1,\ldots,T_m)$, 
\begin{equation*}
\mathfrak{S}_{\cal M} = \{ \mdollarE(\{T_{\tau(d)}\$_d \mid 1\leq d \leq m\}) \mid 
\tau \text{ is permutation of } \{1,\ldots, m\}\}. 
\end{equation*}

This string family coincides with the one introduced in~\cite{BCGR_SAP_CPM2024} (albeit there given with a different definition). 

In this section, we focus on reducing the number of runs of the BWT, commonly referred to as $r$. This is a fundamental parameter, since the BWT is primarily used for compressed text indexes, such as the FM-index~\cite{fm-index-jacm}, the $r$-index~\cite{r-index-jacm}, or the extended $r$-index~\cite{BoucherCLRS21_r,BoucherCLRS24}, whose storage space depends on $r$. Moreover, the value $n/r$, the average runlength of the BWT, is increasingly being used as a parameter describing the input data. This is because it is well known that the more repetitive the input data, the fewer runs the BWT tends to have. 

Therefore, we are interested in the following: 

\begin{definition} \label{def:opt}
Given a string collection ${\cal M}$, we let $r_{\opt}({\cal M}) = \min\{ \rho(L) \mid L \in {\mathfrak S_{\cal M}}\}$, and call 
a string $L \in {\mathfrak S_{\cal M}}$ {\em optimal} if $\rho(L) = r_{\opt}({\cal M})$. When ${\cal M}$ is clear from the context, we just write $r_{\opt}$ for $r_{\opt}({\cal M})$. 
\end{definition}

As we saw in \cref{sec:interesting}, any differences between two transforms $L$ and $L'$ on the same collection ${\cal M}$, in whatever order, necessarily lie within interesting intervals. This ignores the naming of the separators, which is justified, given that no tool outputs the transform with {\em different dollar-symbols}, even if it computes the \mdolBWT\ or \mdolEBWT, which means that internally it handles the dollar-symbols as different symbols.\footnote{Some tools are able to output the indices corresponding to the dollars, such as {\tt BCR} and {\tt BEETL}~\cite{BauerCoxRosoneTCS2013,BCR-tool-BauerCoxRosoneTCS2013,beetl-tool-BauerCoxRosoneTCS2013}, but this is done in a separate file, or the indices can be extracted from the DA or the GSA, such as {\tt eGAP}~\cite{EgidiLMT19_egap,egap-tool-EgidiLMT19_egap}.} Thus, for the sake of compression, it is appropriate to treat the output independently of the names of the dollar-symbols. 

Bentley et al.~\cite{bentley_et_alESA2020opt} gave a linear-time algorithm for minimizing the number of runs but they gave no implementation.
An implementation called {\tt optimalBWT} \cite{optimalBWT-tool-CGLR_DCC2023_opt} was presented by Cenzato et al.~\cite{CGLR_DCC2023_opt} which computes an optimal transform $L$. To understand how the algorithm works, let us consider an interesting interval $[b,e]$. It is clear that within this interval, the number of runs can be minimized if all occurrences of same characters are grouped together. Therefore, if there are $k$ distinct characters in $L[b..e]$, then $k$ is the minimum number of runs achievable in this interval. Now let's say we have grouped all characters into $k$ runs and $L[b..e] = a_1^{r_1}a_2^{r_2}\cdots a_k^{r_k}$. If the previous character $L[b-1] \neq a_1$ then it may be possible to reduce the number of runs by swapping some other run, say $a_i^{r_i}$ with the first run $a_1^{r_1}$,
namely if $L[b-1]=a_i$. Similarly, one may be able to reduce the number of runs by swapping some other run with the last one $a_k^{r_k}$. The problem is that this choice may not be unique, and when there are consecutive interesting intervals then it may not be possible to decide until these interesting intervals end. Bentley et al.\ give an algorithm for making these choices, modeling the problem as what they refer to as a {\em tuple ordering problem}, which can be turned into a shortest path problem on a DAG and solved optimally in linear time. 

Now consider our toy example ${\cal M} = ( {\tt \texttt{CTGA},\texttt{TG},\texttt{GTCC},\texttt{TCA},\texttt{CGACC}, \texttt{CGA}})$. In \cref{tab:heuristics}, we give different transforms from  ${\mathfrak S}_{\cal M}$, with the number of runs between $19$ (top row, input order as given above), and $14$ (bottom row), which is the minimum $r_{\opt}$. In the middle rows we give two heuristics, which we explain next.

\begin{table*}[htbp]
\begin{center}
\scalebox{1}{
\begin{tabular}{l|p{5.9cm}|c|c}
 & transform & number of runs $r$ &  \\
& & &  \\ 
\hline \hline
    $\mdolEBWT({\cal M})$ & {\tt \green{AGCACA}\red{GCG}G\blue{CC}T\orange{TA}\$$_6$\$$_5$\$$_1$T\violet{TC}C\$$_3$\$$_4$G\$$_2$C} & 19 & \\ 
    $\colex({\cal M})$ & {\tt \green{AAACCG}\red{CGG}G\blue{CC}T\orange{AT}\$$_2$\$$_4$\$$_3$T\violet{CT}C\$$_5$\$$_1$G\$$_6$C} & 18 &  \\
    $\plus({\cal M})$ & {\tt \green{AAACCG}\red{GGC}G\blue{CC}T\orange{TA}\$$_2$\$$_5$\$$_1$T\violet{TC}C\$$_4$\$$_3$G\$$_6$C} & 15 &  \\
    $\optimal({\cal M})$ &  {\tt \green{AAAGCC}\red{CGG}G\blue{CC}T\orange{TA}\$$_3$\$$_6$\$$_2$T\violet{TC}C\$$_5$\$$_1$G\$$_4$C} & 14 &  \\
\hline
\end{tabular}

}
\end{center}
\vspace{2mm}
\caption{For the input collection ${\cal M} = ( {\tt \texttt{CTGA},\texttt{TG},\texttt{GTCC},\texttt{TCA},\texttt{CGACC}, \texttt{CGA}})$, we show the resulting transform for the input order, the optimal order, and two heuristics (details see text). Note that in the count of the number of runs, all dollars are considered equal. 
\label{tab:heuristics}
}
\end{table*}

Recall that the colex order (also referred to as {\em reverse lexicographic order (rlo)}) is defined as $S<_{\col} T$ if $S^{\rev}<_{\lex} T^{\rev}$. It is easy to see that if the input strings are in colex order, then every interesting interval will have the minimum number of runs. On our example, the colex results in $18$ runs, one less than the input order. In particular, several existing tools ({\tt BCR}~\cite{BauerCoxRosoneTCS2013,BCR-tool-BauerCoxRosoneTCS2013}, {\tt ropeBWT2}~\cite{ropebwt2-tool-Li14a,Li14a}, {\tt ropeBWT3}~\cite{ropebwt3-tool-ropebwt3,ropebwt3}) have the option to 
output the BWT with respect to the colex order.  
This was already shown in~\cite{CoxBJR_2012_sap} to significantly reduce the number of runs. 

Several other heuristics were introduced in~\cite{BCGR_SAP_CPM2024}, all based on the idea of grouping all occurrences in interesting intervals into one single run and possibly swapping the runs within the interesting interval. The one that performed best was plusBWT, which during the construction 
greedily puts as first run the character immediately before the interesting interval and as last run the character immediately after, if these characters are present. 
In \cref{tab:resExp}, we report experimental results, over benchmarks that are described in Table~\ref{tab:datasets},  that show that both 
of these heuristics get very close to the optBWT. Interestingly, even randomly permuting the runs within interesting intervals results in a slight improvement over the colex order. For details, see~\cite{BCGR_SAP_CPM2024,BauerCoxRosoneTCS2013}.

What is also remarkable is the improvement that can be achieved by the optimalBWT as compared to the input order, namely up to a multiplicative factor of over $31$, on real biological data~\cite{CGLR_DCC2023_opt,BCGR_SAP_CPM2024}, see \cref{tab:resExp}. 
In other words, a resulting data structure built on top of the BWT would be far smaller if the optBWT was used, rather than just any BWT transform. Indeed, in~\cite{CGLR25}, first experimental results on the size of the resulting $r$-index on real datasets are presented. 

Another interesting aspect is the following: in order to reduce the number of runs, a necessary step is grouping all equal characters together within the interesting intervals. The larger the interesting intervals, the fewer the fraction of permutations of the string collection that do this. This implies that the probability that some arbitrary input order---as given by the order in which the collection comes, e.g.\ as downloaded from a database---will group these characters together is very small. This explains why the potential gain from applying the \optimal\ (or, indeed, one of the heuristics given above) is highest on very large sets of very similar short strings: many suffixes will be shared by many input strings, which produces long interesting intervals. (See~\cite{CenzatoL22,CenzatoLiptak_bioinformatics2024} for a more detailed analysis.) 


Now consider the following example, which shows that with the \concat, it can be impossible to minimize the number of runs of the BWT. In other words, on some data sets, whatever the input order, $\concat({\cal M})$ will always have more runs than necessary. 

\begin{example}\label{ex:concat3}
Let ${\cal M} = \{\tt ACA,TGA,GAA\}$, then $t = {\tt acb}$, 
and the output order corresponding to the colexicographic order of the strings is ${\tt bac}$. As it happens, this order yields the minimum number of runs among all strings 
in ${\mathfrak S}_{\cal M}$, with $\colex({\cal M}) =$ $\optimal({\cal M}) =$ {\tt AAAACGG\$AT\$\$}, which has $7$ runs. However, as we saw in \cref{sec:output_order}, the string ${\tt bac}$ does not lie in the image of $\pi_{concat}$, so no permutation of the strings in ${\cal M}$ will yield this order for \concat. In particular, the $\colex({\cal M}) =$ {\tt AAAACGG\$AT\$\$} has $7$ runs, while all concatBWTs have at least $8$: {\tt AAAGACG\$AT\$\$}, {\tt AAACGAG\$AT\$\$}, {\tt AAAAGCG\$AT\$\$}, {\tt AAAGCAG\$AT\$\$}, {\tt AAACAGG\$AT\$\$}. (Here we are using the adapted variant of the \concat, for better comparison.)
\end{example}

\bigskip\bigskip 
\begin{table}[t!]
{\footnotesize
\centering
\renewcommand{\arraystretch}{1.1}
\begin{tabular}{ |l|l|r|r|r|r| } 
 \hline
 \multicolumn{1}{|c|}{\multirow{ 2}{*}{Dataset}}
 & \multicolumn{1}{c|}{\multirow{ 2}{*}{Description}}    & \multicolumn{1}{c|}{\multirow{ 2}{*}{\BWT length}} & \multicolumn{1}{c|}{\textrm{Max }} &  \multicolumn{1}{c|}{Number of} & \multirow{2}{*}{$n/r_{opt}$}  \\ 
 \multicolumn{1}{|c|}{}      & \multicolumn{1}{c|}{}    & \multicolumn{1}{c|}{} & \multicolumn{1}{c|}{\textrm{length}} &  \multicolumn{1}{c|}{sequences} & \\ 
 \hline\hline 
 SRR7494928--30 & {\em Epstein Barr Virus}           & 3,225,480,720        & 101       &   9,648,932 & 79.25 \\
  \hline
 ERR732065--70  & {\em HIV-virus}           & 1,345,713,812        & 150          & 8,912,012  & 116.62 \\ 
 \hline
 SRR12038540    & {\em SARS-CoV-2 RBD}      & 1,690,229,250        & 50                & 33,141,750    &   113.71   \\ 
 \hline  
 ERR022075\_1   & {\em E. Coli str.\ K-12}  & 2,294,730,100        & 100               & 22,720,100  & 32.23 \\ 
 \hline
 SRR059298      & {\em Deformed wing virus} & 2,455,299,082        & 72                & 33,634,234   & 50.75 \\
 \hline 
 SRR065389--90  & {\em C.\ Elegans}         & 14,095,870,474       & 100               & 139,563,074   & 15.30 \\  
 \hline
 SRR2990914\_1  & {\em Sindibis virus}     & 15,957,722,119        & 36          & 431,289,787  & 151.62 \\  
 \hline
 ERR1019034     & {\em H.\ Sapiens}        & 123,506,926,658       & 100         & 1,222,840,858  & 11.37 \\ 
 \hline
 \textit{pdb\_seqres}  & {\em proteins}           & 241,121,574           & 16,181       &   865,773  & 14.33 \\ 
 \hline
\end{tabular}
\vspace{2mm}
\caption{
Features of the datasets included in our experiments. From left to right, we report the dataset code and description, the \BWT\ length, the maximum string length, the number of sequences, and the ratio between the \BWT\ length and the number of runs of the \optimal.
\label{tab:datasets}
}
}
\end{table}

\begin{table}[t!]
\centering
\renewcommand{\arraystretch}{1.1}
{\footnotesize	
\begin{tabular}{|l|r@{\ }|r@{\ }|r@{\ }|r@{\ }|r@{\ }|}
\hline
& & \multicolumn{2}{c|}{\text{Heuristic string orders}} & & \\   \cline{3-4}
\multicolumn{1}{|c|}{\text{Dataset}} & \multicolumn{1}{c|}{\text{inputBWT}} & \multicolumn{1}{c|}{\colex} & \multicolumn{1}{c|}{\text{\plus}} & \multicolumn{1}{c|}{\optimal} & \multicolumn{1}{c|}{$r_{{input}}/r_{{opt}}$ }  \\
\hline\hline 
\text{SRR7494928–30}  & 254,663,327   & 41,730,649   & 41,372,530        & 40,700,607     & \textbf{6.26} \\
    \hline
\text{ERR732065–70}   & 48,727,709     & 11,941,093  & 11,766,827 &   11,539,661  &     \textbf{4.22}  \\
    \hline
  \text{SRR12038540}   & 209,136,502    & 17,026,009  & 15,226,766   &     14,864,523  &     \textbf{14.07}   \\
    \hline
   \text{ERR022075\_1}  & 259,821,570    & 75,846,202  & 74,529,428  &   71,203,469  &  \textbf{3.65}    \\
       \hline
 \text{SRR059298}    & 249,873,376    & 50,495,777  & 49,619,150  &     48,376,632  &  \textbf{5.17}    \\
     \hline
\text{SRR065389–90}   & 2,251,887,226  & 968,098,124     &  954,489,749 & 921,561,895   & \textbf{2.44}   \\
    \hline
\text{SRR2990914\_1}   & 3,313,966,937  & 109,772,697 & 108,466,351  &     105,250,120  &  \textbf{31.49}   \\
    \hline
\text{ERR1019034}    & 23,084,021,291 & 11,312,737,256 & 11,179,873,104  &   10,860,229,434  & \textbf{2.13}  \\
    \hline
\textit{pdb\_seqres}   & 17,971,532     & 16,862,960  & 16,848,496   & 16,829,629      & \textbf{1.07}  \\
\hline
\end{tabular}
}
\vspace{2mm}
\caption{Number of runs of the \mdolEBWT\ (resp. \mdolBWT) with sequences presented in input order (inputBWT) and in two heuristic orders, \colex\ and \plus, compared to the \optimal\ for all datasets in Table~\ref{tab:datasets}. In the rightmost column we report the ratio between the number of runs of the inputBWT ($r_{input}$) and the \optimal\ ($r_{opt}$). Results reported from~\cite{BCGR_SAP_CPM2024}.}
\label{tab:resExp}
\end{table}

\medskip


\section{BWT for string collections in bioinformatics}\label{sec:bioinformatics}

Recent advances in DNA and RNA sequencing technologies led to the explosion of the amount of \emph{in silico} genomic data to be assembled, investigated, stored and compared. In this final section we will overview the applications in bioinformatics that benefit from the efficiency of BWT-based indexing  of string  collections.\\

Both DNA and RNA sequences can be seen as strings over an alphabet of size four: the set of nucleotides that compose them. Furthermore, as we will see below, in all the applications we will mention, sequences are highly repetitive and the different strings of the collection encode redundant data: an ideal target for BWT based tools that are  specifically designed for string collections.\\

When the string collection consists of \emph{raw sequencing data}, that is, a large amount of fragments that directly result from the sequencing process, the most requested tasks can be:

\begin{description}
    \item[Assembling.]  Size and accuracy details have changed across the time and also change from a technology to another, but all sequencing machines are not able to process molecules of arbitrary size.
    As a consequence, the typical procedure is to: first, create a redundant number of copies of the sequence while still \emph{in vitro}; second, cut (this can be done using specific enzymes) the sequences into small enough fragments randomly enough to ensure that distinct copies are cut at different sites; third, perform the actual sequencing that turn the sequence into \emph{in silico} data; fourth, exploiting the redundancy of the original copies and hence fragments overlap, build a layout for the fragments that enables to reconstruct the original sequence.
    This last task is named \emph{Fragment Assembly}, and its efficient solution played a crucial role in the historical human genome project. For decades the problem of fragment assembly has been the \emph{Holy Grail} of algorithmic tasks for bioinformatics, having to deal with millions of DNA fragments, each of a few hundreds of length, that had to be jointly analysed, looking for local similarities among them \cite{DBLP:journals/bioinformatics/SimpsonD10}. 
    
    \item[Mapping.] When the newly sequenced genome belongs to species for which a fully assembled \emph{reference} genome already exists, then one does not need to perform a \emph{de novo} assembly out of the fragments, as the reference can be exploited. In this case, the strings of the collection are, in general, not as many as in the case of the assembly task, because redundancy is not required as much as there. However, a huge amount of strings must be mapped onto the reference genome, dealing with sequencing errors, repetitions inside the genome that make the mapping loci ambiguous, as well as the physiological differences among the query genome and the reference \cite{bowtie,bwa}.
\item[Haplotype Matching.] A Haplotype is a set of genetic markers or DNA variations that are inherited together from a single parent. Haplotype Matching asks for finding long matches between sequences within a given large collection of aligned genetic sequences (that are, indeed, the haplotypes). The task is relevant because these long matches are candidates to be regions that are identical by descent (IBD) from a common ancestor. Even for this task BWT based methods have been designed: the positional BWT 
\cite{Durbin14,DBLP:conf/wabi/SirenGNPD18,DBLP:journals/bioinformatics/SirenGNPD20,ShakyaNZZ21} for linear strings, and more recently the GBWT for Pangenome Graphs
\cite{Sanaullah2025.02.03.634410}.
    \item[Assembly free (and reference free) variant calling.] Variant calling is the fundamental task that has to be performed with a genome, and it consists of detecting individual traits and genetic diversity of the query genome to discover genetic based diseases, perform personalised medicine, as well as to understand evolution (when variants are annotated at population scale). When a reference genome is not available, this task is performed directly on raw data without a preliminary assembling phase, and hence this requires the investigation and comparison of (pairs of) large collections of strings 
\cite{PrezzaPSR20,PrezzaPSR18,DBLP:journals/almob/PrezzaPSR19,Langmead2009}.
    
\end{description}

When, instead, the string collections are multiple genomes, or several RNA isoforms, or belong to possibly many different genomes, we have a (possibly large) set of strings that are as long as billions of letters. In these cases, the downstream analyses tasks are different, and they include:

\begin{description}
    \item[Comparative genomics and variant calling.] Here two or more (whole) genomes must be compared, one of them 
    possibly being \emph{the} reference. Like for the above mentioned tasks, criticalities arise because of the highly repetitive nature of genomes, and by the different nature of variants that one needs to detect: the so called \emph{point mutations} or the \emph{SNPs} that involve single letters, or the short \emph{INDELs} that are the insertion or the deletion of a short fragment, up to the fragment level mutations: the deletions or insertion or duplication of a possibly large DNA fragment, or the translocation (that is, its position in the genome changes), or the copy number variations (fragments that are commonly present in multiple copies show a different number of them in different individuals). Moreover, when a DNA mutation event involves a whole fragment, there is in general a $50\%$ probability that a copy is \emph{inverted}, that is it occurs in the reversed direction.
    The comparison tasks are very complex and cannot be made by means of global alignments due to the fragment level mutations. Hence efficient indices of the strings are a basic toolkit for the various possible algorithmic strategies \cite{BoucherCLRS24,Durbin14}. 
    \item[Transcriptomics.] With the so called new generation sequencing technologies, also \emph{RNA-Seq} became a common task, that is to move to \emph{in silico} also RNA the sequences. In particular, mRNA (messenger RNA) isoforms are very interesting to investigate as they are RNA fragments that encode the information that leads a coding part of the DNA (that is, a gene) to the synthesis of a protein. The transcription of this coding part of DNA into RNA, called \emph{gene expression}, is a very interesting topic for molecular biologists and genetists because the same gene can express in different ways according to the cell it is (say, in the liver rather than in the skin), or according to external conditions (say, in a skin cell whether it is cold or it is being burned). The set of RNA isoforms that are in a certain cell under specific conditions is called the \emph{transcriptome}, and this is a sort of a picture of gene expression. Investigating or comparing transcriptome(s) requires counting the abundance of each kind, being each of them a different concatenation of a subset of the fragments of the gene: again several strings, and again highly redundant, and hence a perfect target for a string collection index based on the BWT \cite{TopHat,BCLRS_SPIRE2021_ebwt,BoucherCLRS24}.
    \item[Metagenomics.] Metagenomics is the (comparative) study of genetic material found in environmental samples (say, into the soil, or in sea water, or in an animal's gut), with the purpose of detecting and investigating (the distribution of) genomes of different species of microorganisms such as bacteria, viruses, and fungi. The ultimate goals of metagenomic is the study of microbial diversity, and the interactions: both mutual and with the ecosystem it belongs to. This requires various tasks that include profiling, assembling, comparing and classifying data consisting of a huge amount of fragments belonging to different individuals and different species. As such, the redundancy will be less than in the other applications; however, indexing these string collections with BWT is still fundamental to make efficient compression and analyses \cite{JaninRC14,GLR2023,BCGR_SAP_CPM2024,GuerriniCGLRT23,GuerriniLR20,salzberg}.  
    \item[Pangenomics.] Using a single genome as \emph{the} reference clearly introduces a bias for the discovery of genomic variations. Therefore, in the literature there came a new challenge that simultaneously deals with, and exploits, the widespread availability of sequencing data: using a \emph{pangenome} rather than a genome. In~\cite{PanGenomeConsortium18}, a pangenome was defined as ``any collection of genomic sequences to be analyzed jointly or to be used as a reference''. In contrast to a \emph{linear} reference, a pangenome reference aims to compactly represent the variation within a population by encoding the commonalities and differences among the underlying sequences. In the literature there are many suggestions for pangenome representations \cite{minigraph,BaaijensBBVPRS22,CarlettiFG0RV19,NatGen2019,PGGB2024,GenRes2017,HumanPG2023,BernardiniPPR20, DBLP:conf/spire/BernardiniPPR17, FMalignGAP_2018,Ohlebusch2019,gramtools_2021} whose construction may benefit from efficient indexing of data as a starting step \cite{LTGPR2020gsufsort,EgidiLMT19_egap,GagieGM21,MunKBGLM20,Oliva0SMKGB21,ShakyaNZZ21,BCLRS_SPIRE2021_ebwt,Sanaullah2025.02.03.634410,CioniGR24}.
    Some of these representations support efficient and accurate analyses tools such as comparison \cite{DBLP:journals/fuin/AlzamelABGIPPR20,DBLP:conf/wabi/AlzamelA0GIPPR18,DBLP:conf/cpm/GaboryMPPRSZ23,fb24,Gabory2025} or alignments \cite{DBLP:conf/biostec/MwanikiGP23,DBLP:journals/siamcomp/BernardiniGPPR22,DBLP:conf/isbra/MwanikiP22,DBLP:conf/icalp/0001GPPR19,cpm18,sopang}, that could be sped up with string collections indices.
    However, by now none of these graph structures has been acknowledged as the standard yet. 
    The computational challenges posed by pangenomes often result in a trade-off between the efficiency and accuracy of the methods and the information content of the chosen representation. In this view, 
    a valid option for certain tasks turns out to be to just index and investigate the collection of genomes, disregarding loci and alignment information and rather choosing speed and low memory usage \cite{phoni,KuhnleMBGLM20,BGKLMM2019_bigBWT,Durbin14}.

\end{description}

\section{Conclusion}\label{sec:conclusion}

In this paper, we studied different methods for constructing the BWT of string collections (as opposed to individual strings). We showed that the methods in use are non-equivalent, in the sense that they produce different transforms, which differ not only in the denomination of separator-symbols (``dollars''). We analyzed precisely where and how these differences occur (``interesting intervals''). The differences between the transforms extend to the number of runs $r$ of the BWT, a central parameter in compressed text indexing. 

We showed that several of these methods are input-order dependent, and that this can be exploited to significantly reduce the number of runs: by up to a factor of over $31$, on real biological data. This is the multiplicative gain of the \optimal\ as compared to computing the transform on the collection in the order it comes in. We also saw that several simple heuristics come close to this optimum. 

On the downside, we saw that one of the most common methods, which we refer to as \concat\ (concatenating the input strings and separating them with the same dollar), has several undesirable properties. First, it cannot produce all possible transforms, and can therefore in general not be used as a method for run minimization. Second, it is not robust with respect to addition or removal of strings from the collection (dynamicity). Finally, it is input-order dependent but in a fairly opaque manner (its order being in essence the BWT of the input order), which makes it difficult to control the output in any way.  

Based on our findings, we make the following suggestions: 

\begin{enumerate}
\item We advise to exercise care when using the method we term \concat, due to the issues listed above. It is a good method because it is simple and efficient, but the user (or programmer) should be aware of its shortcomings. 
\item We believe that the definition of the number of runs of a string collection should be standardized to $r_{\opt}$, the minimum number of runs among all possible separator-based transforms. This number is also fairly easy to approximate with heuristics, the simplest of which is \colex. 
\end{enumerate}


\bibliography{biblio.bib}

\appendix 

\end{document}